\documentclass[12pt]{article}

\usepackage[utf8]{inputenc}
\usepackage{multicol}
\usepackage[scale=0.97]{XCharter}
\usepackage[xcharter, bigdelims, vvarbb, scaled=1.05]{newtxmath}
\usepackage{fullpage}
\usepackage[pdftex,bookmarks,hidelinks]{hyperref}
\usepackage{graphicx}
\usepackage{cleveref}
\usepackage[range-units=brackets, range-phrase=-]{siunitx}[=v2]
\usepackage[svgnames,dvipsnames,x11names,table]{xcolor}
\usepackage{lipsum}
\usepackage[sort&compress,numbers]{natbib}
\usepackage{hyperref}
\usepackage[total={6.5in,9in},top=1in,headsep=0.1in,headheight=1in]{geometry}
\usepackage{aas_macros}

\def\Neff{N_{\rm eff}}

\makeatletter
\g@addto@macro\bfseries{\boldmath}
\makeatother

\definecolor{colorJM}{rgb}{.7,.5,0}

\newcommand\snowmass{
\begin{center}
  \rule[-0.2in]{\hsize}{0.01in}\\
  \rule{\hsize}{0.01in}\\
  \vskip 0.1in
  Submitted to the Proceedings of the US Community Study\\ 
  on the Future of Particle Physics (Snowmass 2021)\\
  \rule{\hsize}{0.01in}\\
  \rule[+0.2in]{\hsize}{0.01in}\\[-2em]
\end{center}
}

\usepackage[firstpage=true]{background}
\backgroundsetup{contents={\parbox{6.5in}{\snowmass}}, scale=1,placement=top,opacity=1,color=black,position={3.25in,1.2in}}

\usepackage{fancyhdr}
\fancypagestyle{plain}{%
  \fancyhf{}%
  \fancyhead[C]{}
  \fancyfoot[C]{\thepage}
}

\fancypagestyle{empty}{%
  \fancyhf{}%
  \fancyhead[C]{{\it The Physics of Light Relics}}
  \fancyfoot[C]{\thepage}
}
\title{The Physics of Light Relics}
\date{}

\usepackage{authblk}

\author[1]{Cora Dvorkin\footnote{cdvorkin@g.harvard.edu}}
\affil[1]{Department of Physics, Harvard University, 17 Oxford Street, Cambridge, MA 02138, USA}
\author[2]{Joel Meyers\footnote{jrmeyers@smu.edu}}
\affil[2]{Department of Physics, Southern Methodist University, Dallas, TX 75275, USA}

\author[3]{Peter Adshead}
\affil[3]{Illinois Center for Advanced Studies of the Universe, and Department of Physics, University of Illinois at Urbana-Champaign, Urbana, Illinois, 61801, USA}
\author[4]{Mustafa Amin}
\affil[4]{Department of Physics and Astronomy, Rice University, Houston, TX 77005, USA}
\author[1]{Carlos A. Arg\"uelles}
\author[5,6]{Thejs Brinckmann}
\affil[5]{Dipartimento di Fisica e Scienze della Terra, Universit\`a di Ferrara, Polo Scientifico e Tecnologico - Edificio C Via Saragat~1, 44122 Ferrara, Italy}
\affil[6]{Istituto Nazionale di Fisica Nucleare (INFN), Sezione di Ferrara, Via Giuseppe Saragat 1, 44122 Ferrara, Italy}
\author[7]{Emanuele Castorina}
\affil[7]{Dipartimento di Fisica ‘Aldo Pontremoli’, Universita' degli Studi di Milano, Via Celoria 16, 20133 Milan, Italy}
\author[8]{Timothy Cohen}
\affil[8]{Institute for Fundamental Science, Department of Physics, University of Oregon, Eugene, OR 97403, USA}
\author[9]{Nathaniel Craig}
\affil[9]{Department of Physics, University of California, Santa Barbara, CA 93106, USA}
\author[10]{David Curtin}
\affil[10]{Department of Physics, University of Toronto, Canada}
\author[11]{Francis-Yan Cyr-Racine}
\affil[11]{Department of Physics and Astronomy, University of New Mexico, Albuquerque, NM 87106, USA}
\author[12]{Peizhi Du}
\affil[12]{C. N. Yang Institute for Theoretical Physics,
Stony Brook University, Stony Brook, NY 11794, USA}
\author[13]{Lloyd Knox}
\affil[13]{Physics Department, University of California, Davis, CA 95616, USA}
\author[14,15]{Bohua Li}
\affil[14]{School of Physical Science and Technology, Guangxi University, Nanning, 530004, China}
\affil[15]{Department of Astronomy, Tsinghua University, Beijing, 100084, China}
\author[16]{Marilena Loverde}
\affil[16]{Department of Physics, University of Washington, Seattle, WA 98195, USA}
\author[3]{Kaloian Lozanov}
\author[17]{Julian B.~Mu\~{n}oz}
\affil[17]{Center for Astrophysics \textbar{}  Harvard \& Smithsonian, 60 Garden Street, Cambridge, MA 02138, USA}
\author[18]{Katelin Schutz}
\affil[18]{Department of Physics \& McGill Space Institute, McGill University, Montr\'{e}al, QC H3A 2T8, Canada}
\author[19]{Paul Shapiro}
\affil[19]{Department of Astronomy and Texas Cosmology Center, The University of Texas at Austin, 2515 Speedway C1400, Austin, Texas 78712, USA}
\author[20,21]{Benjamin Wallisch}
\affil[20]{School of Natural Sciences, Institute for Advanced Study, Princeton, NJ~08540, USA}
\affil[21]{Department of Physics, University of California San Diego, La Jolla, CA~92093, USA}
\author[16]{Zachary J. Weiner}
\author[22]{Weishuang Linda Xu}
\affil[22]{Berkeley Center for Theoretical Physics, South Hall Rd, Berkeley, CA 94720, USA}

\begin{document}

\maketitle
\begin{abstract}
Many well-motivated extensions of the Standard Model predict the existence of new light species that may have been produced in the early universe. 
Prominent examples include axions, sterile neutrinos, gravitinos, dark photons, and more.
The gravitational influence of light relics leaves imprints in the cosmic microwave background fluctuations, the large-scale structure of the universe and the primordial element abundances. In this paper, we detail the physics of cosmological light relics, and describe how measurements of their relic density and mass serve as probes of physics beyond the Standard Model. 
A measurement of the light relic density at the precision of upcoming cosmological surveys will point the way toward new physics or severely constrain the range of viable extensions to the Standard Model.
\end{abstract}

\tableofcontents

\section{Introduction}

Cosmological observations enable precise measurements of the densities and interactions of the various constituents of the universe.  These measurements have provided firm evidence for the existence of dark energy and non-baryonic dark matter (DM), and the quest to understand the nature of these components remains among the most pressing challenges in fundamental physics.  Upcoming cosmological surveys will drastically improve the precision with which we measure the radiation density of the universe.  A precise measurement of the radiation density may reveal the existence of new dark radiation, or it may place broad constraints on the types of physics that operate in the dark sector.

Many well-motivated extensions of the Standard Model (SM) predict the existence of new light species.  As such, the hunt for new contributions to the radiation density is much more than just a speculative blind search.  As will be described below, models that can explain the physics of the dark sector, address the strong CP problem, solve the hierarchy problem, and account for short baseline neutrino anomalies all contain new light degrees of freedom that can be detected or severely constrained with upcoming cosmological observations.

It is convenient to discuss the light relic density in terms of the effective number of neutrino species $\Neff$ defined as
\begin{equation}
    \rho_r = \rho_\gamma \left(1+\frac{7}{8}\left(\frac{4}{11}\right)^{4/3} \Neff \right) \, ,
\end{equation}
where $\rho_r$ is the total radiation energy density and $\rho_\gamma$ is the energy density of photons.  This definition is chosen such that $\Neff$ would count the number of flavors of neutrino species had SM neutrinos decoupled instantaneously prior to electron-positron annihilation.  In reality, neutrino decoupling was not instantaneous, and the Standard Model prediction is $N_{\rm eff}^{\rm SM} = 3.044(1)$~\cite{EscuderoAbenza:2020cmq,Akita:2020szl,Froustey:2020mcq,Bennett:2020zkv}.
As can be seen from this definition, the term `light relics' does not imply a strict cut-off in mass scale, but rather refers to all species that are relativistic during the epoch relevant for a given observable.  For example, thermal relics with masses below about an \si{MeV} contribute to the light relic density inferred from primordial light element abundances, while only those with masses below a few \si{eV} will be relativistic during recombination and contribute to the light relic density inferred from the cosmic microwave background (CMB) and large-scale structure (LSS).  Non-thermal relics could be much more massive and still contribute to the light relic density, so long as they have a phase space density that makes them relativistic during the relevant period.

The well-understood thermal history of the universe within the Standard Model allows for the identification of compelling thresholds in the search for new light species.  Any model containing new light degrees of freedom that had ever been in thermal equilibrium with the Standard Model plasma in the early universe will predict a value of $\Neff$ that exceeds that in the Standard Model, with $\Delta \Neff = \Neff-N_{\rm eff}^{\rm SM} \geq 0.027$ for each new light degree of freedom.  This places the cosmological signatures of many models of new physics within reach of the upcoming generation of cosmological surveys. 

Even models that do not involve new light states can predict interesting signatures in the light relic density.  Modified thermal histories can alter the Standard Model prediction of $\Neff$, including the possibility of $\Delta \Neff<0$.  The stochastic gravitational wave background contributes to the radiation density of the universe, and measurements of $\Neff$ thereby serve as an integral constraint on the gravitational wave spectrum.

Furthermore, light relics may have a non-zero mass, and this could leave observable imprints in the LSS of the universe. 
A light relic that is non-relativistic today will have a contribution to the DM density of the universe, and because of their non-zero temperatures we can distinguish them from the majority of cold dark matter.
We will show how this opens up a new route for finding physics beyond the Standard Model with cosmological observations.

In the sections below we will explore the physics of cosmological light relics and describe how measurements of the light relic density and mass serve as a broad and useful probe of physics beyond the Standard Model.

\section{Light thermal relics}

The early universe was filled with a hot dense plasma and underwent a very rapid, radiation-dominated expansion.  The conditions during this early phase were sufficiently extreme that particle-antiparticle pairs of all sorts were rapidly produced and annihilated in frequent energetic collisions.
Any species with sufficient couplings to Standard Model degrees of freedom would have been abundantly produced in the hot dense conditions of the early universe.  Particles which were once in thermal equilibrium and are stable on cosmological time scales persist as thermal relics of the hot Big Bang.
Massive thermal relics became non-relativistic at early times and provide candidates for cold dark matter.
Light thermal relics remained relativistic throughout the early universe, contributing to the radiation density during this epoch.

\subsection{Thermal freeze-out of light relics}
\label{sec:freezeout}
The relic density of any light thermal relic can be straightforwardly computed simply through considerations of comoving entropy conservation.
After a light species decouples from the thermal plasma, it will maintain its relativistic distribution function.
The energy density of decoupled species is diluted compared to that of photons by the disappearance of species in thermal equilibrium as the temperature drops below their masses.
The annihilation of these states causes their entropy density to be passed to that of the plasma, eventually winding up in the photons that make up the CMB.
By counting the degrees of freedom that annihilate after the decoupling of a light thermal relic species, it is straightforward to compute its energy density compared to that of photons.

\begin{figure}[t]
	\centering
	\includegraphics{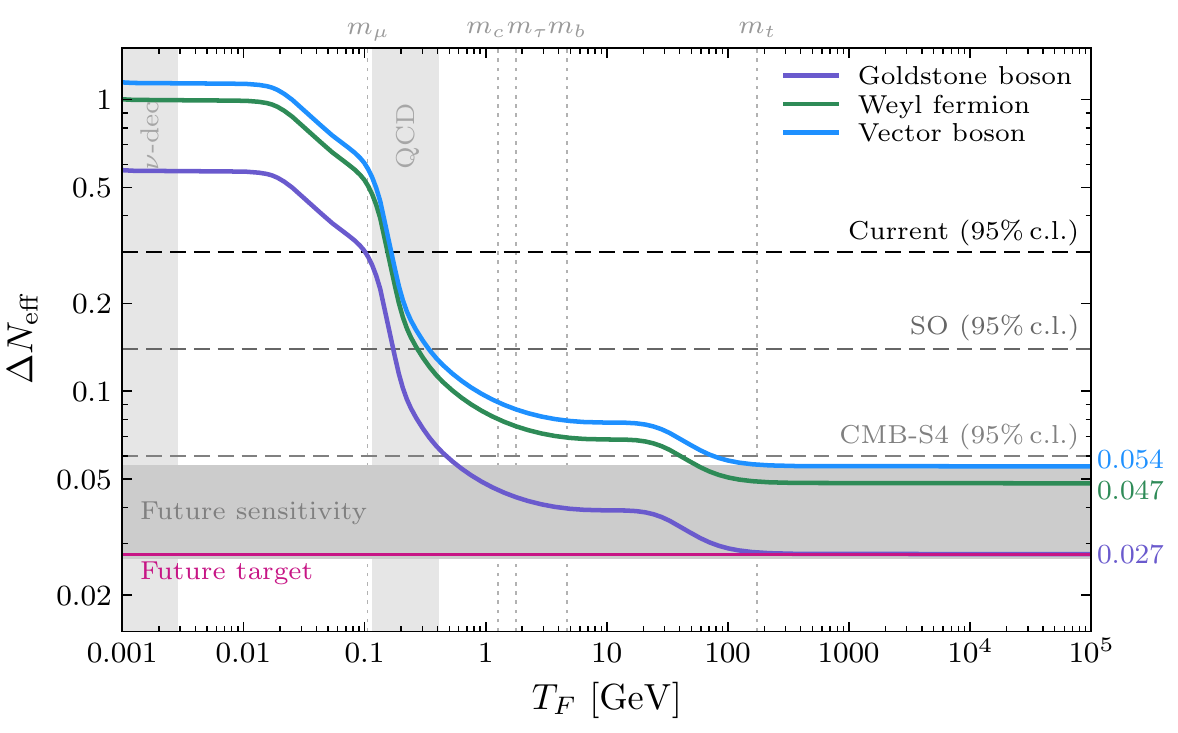}
	\caption{Contributions of a single massless particle, which decoupled from the Standard Model at a freeze-out temperature~$T_F$, to the effective number of relativistic species, $\Neff = N_\mathrm{eff}^\mathrm{SM} + \Delta\Neff$, with the Standard Model expectation $N_\mathrm{eff}^\mathrm{SM} = 3.044$ from neutrinos~(adapted from~\cite{Wallisch:2018rzj, Green:2019glg}). The purple, green and blue lines show the contributions for a real scalar, Weyl fermion and vector boson, respectively. The drop in~$\Delta\Neff$ by about one order of magnitude around $T_F \sim \SI{150}{MeV}$ is due to the QCD~phase transition, which is denoted by a vertical gray band, as is neutrino decoupling. The dashed lines indicate the current bound on~$\Delta\Neff$ at 95\%~C.L.\ from {\it Planck}~2018 and BAO~data~\cite{Planck:2018vyg}, and the anticipated sensitivity of the Simons Observatory~(SO)~\cite{SimonsObservatory:2018koc} and \mbox{CMB-S4}~\cite{Abazajian:2019eic,Snowmass2021:CMBS4} as examples for upcoming and next-generation CMB experiments~(see also~\cite{Hanany:2019lle}, for instance). The horizontal gray band illustrates the future sensitivity that might potentially be achieved with a combination of cosmological surveys of the~CMB and large-scale structure, such as CMB-HD~\cite{Sehgal:2019ewc}, MegaMapper~\cite{Schlegel:2019eqc} and PUMA~\cite{PUMA:2019jwd}, cf.~\cite{Baumann:2017gkg, Sailer:2021yzm, MoradinezhadDizgah:2021upg}. The displayed values on the right are the observational thresholds for particles with different spins and arbitrarily large decoupling temperature. We refer to~\cite{Wallisch:2018rzj, Green:2019glg} for additional details.}
	\label{fig:deltaNeff_freezeout}
\end{figure}

The contribution of a light thermal relic species $X$ to the radiation density is
\begin{equation}
    \Delta \Neff = \frac{4}{7}g_{X,\star} \left(\frac{T_X}{T_\nu}\right)^4 = \frac{4}{7} g_{X,\star} \left(\frac{43}{4g_\star(T_F)}\right)^{4/3} \ ,
    \label{eq:Delta_Neff}
\end{equation}
where $g_{\chi,\star}$ is the effective number of degrees of freedom of $X$, including an additional factor of $7/8$ for fermionic particles, $T_\nu$ is the temperature of the cosmic neutrino background, given by $T_\nu = (4/11)^{1/3}T_\gamma$, where $T_\gamma$ is the temperature of the cosmic microwave background, and $g_\star(T_F)$ is the effective number of relativistic degrees of freedom in thermal equilibrium at the temperature $T_F$ at which $X$ decouples from the plasma. Since we know the particle content of the Standard Model and therefore $g_\star(T)$, we can predict the relic density of any light thermal relic from its spin and its freeze-out temperature $T_F$; see Figure~\ref{fig:deltaNeff_freezeout}. The freeze-out temperature of a given relic is approximately given by the temperature at which the rate of interactions with the Standard Model, $\Gamma$, drops below the expansion rate $H$, which during the radiation-dominated era is given by $H^2 = \frac{\pi^2}{90}g_\star(T)\frac{T^4}{M_\mathrm{pl}^2}$.  

Let us consider a species that has an interaction rate with the Standard Model that scales as $\Gamma \sim \lambda^2 T^{2n+1}$, where $\lambda$ is a coupling constant with units of $[\mathrm{Energy}]^{-n}$, as is expected when the interaction is mediated by an operator of dimension greater than four. The ratio of interaction rate to expansion rate is then $\Gamma/H \sim \lambda^2 T^{2n-1}M_\mathrm{pl}$, and a given species will be in equilibrium when $\lambda^2 \gg M_\mathrm{pl}^{-1}T^{-2n+1}$.  For $n \geq 1$, this implies that the minimum coupling necessary for a species to be in equilibrium scales as an inverse power of the temperature.

The extremely high temperatures achieved in the early universe suggest that even very feebly interacting particles would have been in equilibrium at sufficiently early times.  Furthermore, at temperatures above the QCD phase transition, $g_\star$ changes relatively slowly with temperature, and thus even small improvements in the measurement of $\Neff$ translate into drastically improved reach in freeze-out temperature and thus the couplings of new light species.  The relic density of any new light species that was ever in thermal equilibrium with the SM~plasma is given by $\Delta \Neff \geq 0.027$ (unless there are changes to the thermal history). Measurements of the light relic density therefore serve as an extremely broad probe of new physics.

\subsubsection*{The Cosmic Neutrino Background}
Cosmic neutrinos provide a useful and familiar example of the freeze-out of light thermal relics within the Standard Model~\cite{Snowmass2021:CosmoLabNeutrinos}.  Neutrinos interact with charged leptons through the weak nuclear force, and decoupled from the thermal plasma when the temperature dropped below about $\SI{1}{MeV}$.  Shortly after neutrino decoupling, electrons and positrons annihilated, causing photons to be heated relative to neutrinos, such that $T_\nu/T_\gamma = (4/11)^{1/3}$ after electron-positron annihilation.  Cosmic neutrinos continued to make a significant contribution to the radiation energy density, impacting the expansion rate and the evolution of cosmological fluctuations through their gravitational influence.

\subsection{Rethermalization via Standard Model fermions}
\label{sec:rethermalization}
New light particles can also be out of equilibrium at high temperatures in the early universe, but come into thermal equilibrium at temperatures below the electroweak symmetry breaking~(EWSB) scale through interactions with one or more (massive) SM~fermions. This is because the production rate of these particles below the scale of~EWSB and above the masses of the respective fermions scales as $\Gamma_n \propto T^{2n+1}$ with $n\leq0$, which is weaker than the temperature dependence of the expansion rate, $H \propto T^2$, in contrast to the freeze-out scenario discussed above. In consequence, this leads to a more complicated thermal evolution. Once these particles thermalize with the~SM, they will contribute to~$\Neff$ at an observable level. In fact, this would be easier to detect since the contribution to~$\Neff$ in this rethermalization scenario is larger than their equivalent freeze-out contribution. At the same time, the absence of a detection would allow us to place direct constraints on the interaction strength between the light particles and the SM~fermions~\cite{Baumann:2016wac}.\medskip

To be more specific, we can consider the couplings of SM~fermions to pseudo-Nambu-Goldstone bosons~(pNGBs), which are scalar particles that arise as a consequence of the breaking of (approximate) global symmetries, such as axions~(shift symmetry)~\cite{Peccei:1977hh, Weinberg:1977ma, Wilczek:1977pj, Hook:2018dlk}, familons (flavor symmetry)~\cite{Davidson:1981zd, Wilczek:1982rv, Reiss:1982sq, Feng:1997tn}, majorons~(neutrino masses)~\cite{Chikashige:1980ui, Chacko:2003dt} or the gravitino~(supersymmetry). Of general interest is the dimension-5 derivative coupling of a pNGB~$\phi$ to the SM~axial vector current,~$\partial_\mu \phi\, \bar{\psi} \gamma^\mu \gamma^5 \psi$, for instance. In this case, the approximate chiral symmetry of the fermions makes this interaction effectively marginal below the EWSB~scale resulting into an interaction rate with~$n=0$, $\Gamma_0 \propto T$. In consequence, the out-of-equilbrium~pNGB recouples to the SM~bath at low temperatures. However, this is only true above the mass of the SM~fermion since the production rate will again become negligibly small once the number density of the fermion is sufficiently Boltzmann suppressed, which implies that the effective decoupling temperature is below the associated fermion mass, resulting in the corresponding contribution to~$\Neff$.

\begin{figure}[t]
	\centering
	\includegraphics{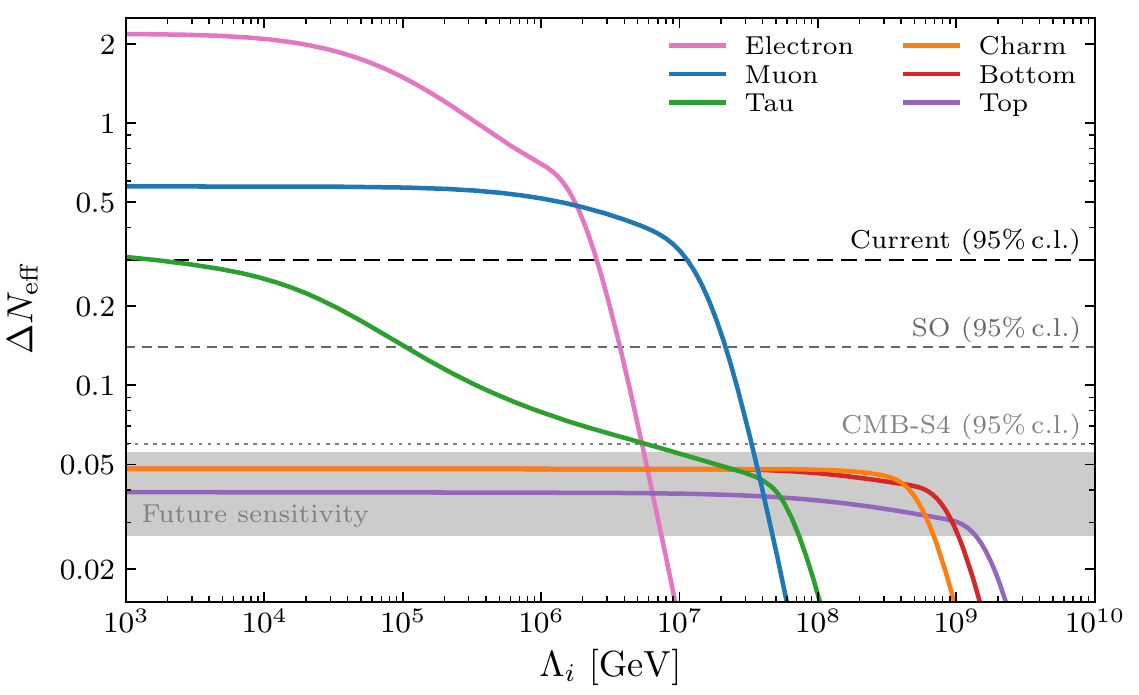}
	\caption{Contribution to the radiation density as parametrized by~$\Delta\Neff$ as a function of the axion interaction strength~$\Lambda_i$ to different Standard Model fermions~$\psi_i$ for the dimension-5 derivative coupling to the SM~axial vector current,~$\mathcal{L} \supset -\Lambda_i^{-1}\partial_\mu \phi\, \bar{\psi}_i \gamma^\mu \gamma^5 \psi_i$~(taken from~\cite{Green:2021hjh}). The displayed values for the bottom and charm couplings are conservative and may be~(significantly) larger. The horizontal dashed lines and gray band are the same as in Fig.~\ref{fig:deltaNeff_freezeout}. From this figure, current and future constraints on~$\Neff$ can be directly translated into the equivalent bounds on~$\Lambda_i$ for any couplings to matter. We refer to~\cite{Green:2021hjh} for a detailed discussion.}
	\label{fig:deltaNeff_axionMatterCouplings}
\end{figure}

To avoid this large energy density of~pNGBs requires that the recoupling temperature is smaller than this effective decoupling temperature so that the interaction rate is already Boltzmann suppressed when the~pNGBs would rethermalize. This requirement can be expressed as a bound on the interaction strength coupling the~pNGBs to the SM~fermions. Although these constraints are usually weaker than the freeze-out bounds, they have the advantage that they do not make any assumptions about the reheating temperature~(as long as reheating occurs above the relevant SM~fermion mass). As a consequence, both current constraints on~$\Neff$ from {\it Planck} and~BBN, and in particular future measurements of~$\Delta\Neff$ at the projected sensitivity of~CMB-S4 set interesting constraints on the couplings of some SM~fermions to axions and other~pNGBs, see e.g.~\cite{Baumann:2016wac, Ferreira:2018vjj, DEramo:2018vss, Arias-Aragon:2020shv, Ghosh:2020vti, Ferreira:2020bpb, Dror:2021nyr, Green:2021hjh}. This is illustrated in Figure~\ref{fig:deltaNeff_axionMatterCouplings}. %
In particular, the cosmological constraints on axion-muon and axion-tau couplings are currently complementary to astrophysical constraints from supernova~1987A~(cf.~\cite{Bollig:2020xdr, Croon:2020lrf}) and white dwarf cooling, respectively, and will be competitive or even supersede these bounds with future $\Neff$~measurements~\cite{Green:2021hjh}. While the derivation of precise constraints on the couplings of axions to the heavy quarks depends on the strong-coupling regime of~QCD, upcoming CMB~and LSS~experiments will also be sensitive to these interactions. More generally, the required modeling, physical understanding and the environment in the early universe is quite different to the interiors of stars which means that cosmological probes can be an important complementary test of these pNGB-matter couplings (see e.g.~\cite{DeRocco:2020xdt}).\medskip

\section{Fluid-like light relics} 

Cosmological observables are sensitive to more than just the total energy density of light relics.  Perturbations in the light relic density affect photon, baryon, and dark matter perturbations due to their gravitational interactions, which can in turn be observed in CMB and large-scale structure surveys.  This sensitivity to perturbations in the density of light relics provides a cosmological window into the interactions of light relic species.

Light relic particles that have no significant non-gravitational interactions propagate at nearly the speed of light in the early universe.  Perturbations to the light relic density of such free-streaming radiation therefore propagate at a speed that exceeds the sound speed of the photon-baryon plasma, leading to anisotropic stress.  The gravitational influence of the free-streaming radiation beyond the sound horizon of the plasma leads to a characteristic phase shift in the spectrum of acoustic oscillations~\cite{Bashinsky:2003tk,Baumann:2015rya,Baumann:2017lmt}.  The phase shift due to free-streaming cosmic neutrinos has been measured in the CMB~\cite{Follin:2015hya} and in baryon acoustic oscillations~\cite{Baumann:2019keh}. For a more detailed discussion, we refer to the dedicated Snowmass~2021 White Paper on neutrinos in cosmology (and the laboratory)~\cite{Snowmass2021:CosmoLabNeutrinos}.

Light relics exhibiting significant interactions may instead behave like a fluid.  Perturbations to the density of fluid-like light relics propagate with a speed that does not exceed the sound speed of the photon-baryon plasma, and therefore do not lead to a phase shift.  Fluid-like light relics may arise in models with neutrino self-interactions, neutrino-dark sector interactions, or dark radiation self-interactions~\cite{Bell:2005dr,Friedland:2007vv,Cyr-Racine:2013jua,Oldengott:2014qra,Kreisch:2019yzn,Wilkinson:2014ksa,Escudero:2015yka,Buen-Abad:2015ova,Chacko:2016kgg,Buen-Abad:2017gxg,Brust:2017nmv,Choi:2018gho}.

While both free-streaming and fluid-like light relics contribute to the radiation energy density and, therefore, have identical effects on the expansion rate, the behavior of their perturbations is different, allowing them to be distinguished in observations. Current CMB measurements have placed constraints on the amount of energy density in free-streaming and fluid-like light relics. We list the limits from {\it Planck} 2018 TT,TE,EE+BAO data sets~\cite{Planck:2019nip} on different models in Table~\ref{tab:N_eff_constraints}. To avoid confusion, we denote the energy density of free-streaming light relics to be $N_{\rm eff, fs}$ while that of fluid-like relics to be $N_{\rm eff, fld}$. If we fix the contribution from SM neutrinos to be $N_{\rm eff, fs}=3.046$, the constraints are $\Delta N_{\rm eff,fs}<0.28$  and  $\Delta N_{\rm eff,fld}<0.38$ at $95\%$ C.L. (see first and second models in Table~\ref{tab:N_eff_constraints}). In general, CMB data allows larger $\Delta N_{\rm eff,fld}$ due to the physical effect of fluid-like light relics (or lack thereof) discussed above.

\begin{table}[t]
\renewcommand{\arraystretch}{1.2}
\centering
 \begin{tabular}{|c|c|c|}
 \hline
 Model & Constraint from {\it Planck} TT,TE,EE+BAO & Reference\\
 \hline
  $N_{\rm eff,fs}=3.046+ \Delta N_{\rm eff,fs}$& $\Delta N_{\rm eff,fs}<0.28$ & \cite{Planck:2018vyg}\\
  
   \rowcolor[HTML]{EFEFEF} $N_{\rm eff,fs}=3.046$, $\Delta N_{\rm eff,fld}$ & $\Delta N_{\rm eff,fld}<0.38$ & \cite{Blinov:2020hmc}\\
   
   $N_{\rm eff,fs}+ N_{\rm eff,fld}=3.046$ &$N_{\rm eff,fld}<0.6$ & \cite{Blinov:2020hmc}\\
   
   \rowcolor[HTML]{EFEFEF} $N_{\rm eff,fs}$, $N_{\rm eff,fld}$ & $N_{\rm eff,fld}<0.6$ & \cite{Blinov:2020hmc} \\
   
   $N_{\rm eff,fs}+ N_{\rm eff,fld}=3.046$, $\sum m_\nu$ &$N_{\rm eff,fld}<0.5$ & \cite{Brinckmann:2020bcn} \\
   
   \rowcolor[HTML]{EFEFEF} $N_{\rm eff,fs}$, $N_{\rm eff,fld}$, $\sum m_\nu$ & $N_{\rm eff,fld}<0.51$ & \cite{Brinckmann:2020bcn} \\
 \hline
\end{tabular}
\caption{
The $95\%$ C.L. constraints on $N_{\rm eff,fs}$ and $N_{\rm eff,fld}$ from {\it Planck} 2018 TT,TE,EE+BAO data sets. SM neutrinos are assumed to be massless for the first four models and massive for the last two.  Here, we consider six cases: (1) $N_{\rm eff,fs}=3.046+\Delta N_{\rm eff,fs}$, (2)  $N_{\rm eff,fs}=3.046$ with additional $\Delta N_{\rm eff,fld}$, (3) the sum of $N_{\rm eff,fs}$ and $N_{\rm eff,fld}$ is fixed to $N_{\rm eff,fs}+ N_{\rm eff,fld}=3.046$,  (4) both $N_{\rm eff,fs}$ and $N_{\rm eff,fld}$ are allowed to vary, and (5) and (6) are the same as (3) and (4), but with the neutrino mass sum being varied.}
\label{tab:N_eff_constraints}
\end{table}

It is also possible in many scenarios beyond the Standard Model (BSM) that light relics may have additional features that are not captured in the two cases above. For example, the rate of interactions that keep light relics in equilibrium can have a certain time dependence, which might drop below the Hubble rate at an early time or a later time. Depending on the type of interactions, light relics may be fluid-like in the early universe and start to free stream later (the decoupling case), or the other way around (the recoupling case), cf.~Sections~\ref{sec:freezeout} and~\ref{sec:rethermalization}. Therefore, the effects of these types of light relics on the CMB power spectra should be a hybrid of the free-streaming and fluid-like cases. Moreover, due to different properties before and after the transition, decoupling/recoupling light relics leave distinct features on the spectrum that depend on the transition time. Interestingly, models with decoupling~\cite{Kreisch:2019yzn,Park:2019ibn,Ghosh:2019tab,Das:2020xke} and recoupling neutrinos~\cite{Escudero:2019gvw,EscuderoAbenza:2020egd,Escudero:2021rfi,Aloni:2021eaq} are proposed to resolve the $H_0$ tension.  These two cases can arise from majoron models where SM neutrinos couple to a scalar majoron. Depending on the mass of the majoron, neutrinos can have decoupling (recoupling) features if the majoron is heavy (light). While the decoupling case is disfavored by other terrestrial and cosmological constraints~\cite{Blinov:2019gcj,Lyu:2020lps,Deppisch:2020sqh,Brinckmann:2020bcn}, the recoupling neutrino model with an \si{eV}-scale majoron may help mitigate the $H_0$ tension~\cite{Schoneberg:2021qvd}.

\section{Light but Massive Relics - {\it LiMR}s}

\subsection{Motivation}

Light relics appear from new physics at a broad range of energy scales, and  while we commonly assume that the relics themselves are massless, this need not be exactly true.
Light relics with non-zero masses can give rise to interesting cosmological signatures.
A light relic that is non-relativistic today will contribute to the dark-matter density.
However, their non-zero temperatures will allow us to distinguish them from the majority of cold dark matter, which provides us with a new avenue to find light relics in cosmology, as well as to distinguish them from each other and from other cosmological uncertainties.

The best-known example of a light (but massive) relic, or LiMR, are neutrinos.
These particles decoupled while relativistic (at $T\sim1-10$ \si{MeV}~\cite{Mangano:2005cc}), so they keep a large cosmic abundance (with a number density $n_\nu\sim 10^2\,\rm cm^{-3}$ today).
Their non-zero masses, while small enough to be unresolved by current laboratory experiments~\cite{Aker:2019uuj}, can be tightly constrained by cosmological datasets~\cite{Lesgourgues:2006nd}.
Other examples of LiMRs that may populate our universe are massive gauge vectors (i.e.,~dark photons~\cite{Vogel:2013raa,Ackerman:2008kmp}), axions~\cite{Peccei:1977hh, Weinberg:1977ma, Wilczek:1977pj, Arvanitaki:2009fg}, and the gravitino~\cite{Weinberg:1982zq,Feng:2010ij}.
We will return to the different relic examples, and their relation to neutrinos in Section~\ref{sec:LiMR_Mnu}.

\subsection{The effect of relic masses}

Light relics decouple from the SM bath while relativistic, so they retain their original (Fermi-Dirac or Bose-Einstein, assuming thermal equilibrium) phase-space distribution.
Thus, their temperature scales linearly as  
$T_X(z) = T_X^{(0)}(1+z)$ with redshift $z$, and their momentum is $p_X\approx 3 T_X$.
For the common assumption of massless relics, their effect can be fully absorbed into a change to the effective number $N_{\rm eff}$ of neutrino species (see above), which parametrizes extra contributions to cosmic radiation.  
However, massive LiMRs can transition to be non-relativistic when $p_X(z)\approx m_X$, and change from contributing to the radiation energy budget ($\Delta N_{\rm eff}$) to the matter content ($\Omega_M$) of the universe.
Unlike the majority of DM---which is cold---LiMRs have significant velocities due to their temperature, which impedes their clustering beyond a characteristic free-streaming scale~\cite{Lesgourgues:2006nd}. 
Therefore, LiMRs (like neutrinos) behave as a type of hot DM, impacting the growth of matter fluctuations and thus the observable LSS of the universe.
This provides us with an interesting target for cosmological observations.

We can find the scales at which a relic affects the growth of structure by calculating the (comoving) free-streaming scale~\cite{Munoz:2018ajr,Ali-Haimoud:2012fzp}
\begin{equation}
    k_{\rm fs} \approx \frac{0.1\,h\rm Mpc^{-1}}{\sqrt{1+z}} \frac{1.95\,\rm K}{T_X} \frac{m_X}{0.1\,\rm eV}
\end{equation}
as a function of its mass $m_X$ and temperature $T_X$, for a fiducial {\it Planck} 2018 cosmology.
Likewise, the size of the effect of a LiMR in the matter clustering will be given by its cosmic abundance, given by
\begin{equation}
    \omega_X = \Omega_X\,h^2 \approx \frac{m_X}{93.14\,\rm eV} \left( \frac{T_X}{1.95\,\rm K} \right)^3
\end{equation}
in terms of that of neutrinos~\cite{Viel:2005qj,Lesgourgues:2006nd}.
By studying at which location ($k_{\rm fs}$)  the suppression in power due to a LiMR is centered, and how deep it is (which is proportional to $\omega_X$), we can find not only whether there is any cosmic LiMR, but also its particle properties, mass $m_X$ and temperature $T_X$~\cite{Xu:2021rwg}.

\subsection{Current constraints}

Our most straightforward measurements of radiation energy are anchored at the epochs of Big Bang Nucleosynthesis (BBN) and recombination.

\begin{figure}
    \centering
    \includegraphics[width=0.6\textwidth]{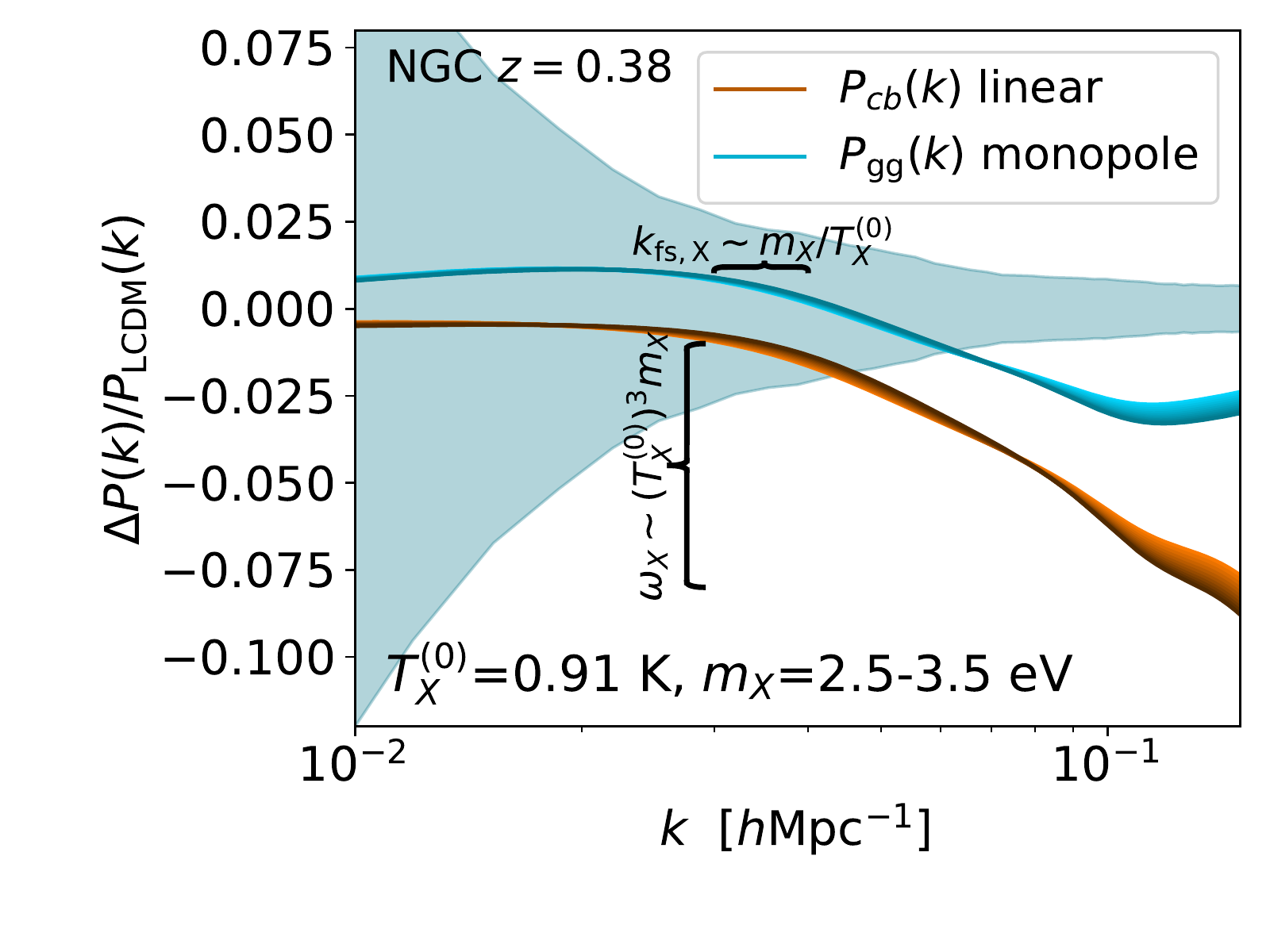}
    \caption{The effect of a single species of Weyl fermion relic at $T_X^{(0)}=0.91$~\si{K} and various masses 2.5 (lightest) to 3.5 (darkest) \si{eV}, on the linear CDM+baryon (orange) and galaxy-galaxy monopole (blue) power spectra. The shaded region is taken to be representative of current experimental shot noise, specifically that of the BOSS DR12 low-redshift, North Galactic Cap catalogue.}
    \label{fig:pk_limr}
\end{figure}

If the relic is non-relativistic at present, then an additional avenue for observation might be found in the large-scale structure of the universe, where the free-streaming of the species at small-scales will impede the formation of structure therein. 
Figure~\ref{fig:pk_limr} illustrates this suppression effect in the linear-theory matter power spectrum (orange) and in the galaxy power spectrum (after perturbative corrections have been included) (blue). 
As the relics redshift into becoming non-relativistic, a joint analysis of cosmological data sets at different epochs (for instance CMB, LSS, and weak lensing) is important to maximally search for these degrees of freedom.

\begin{figure}[t!]
    \centering
    \includegraphics[width=0.6\textwidth]{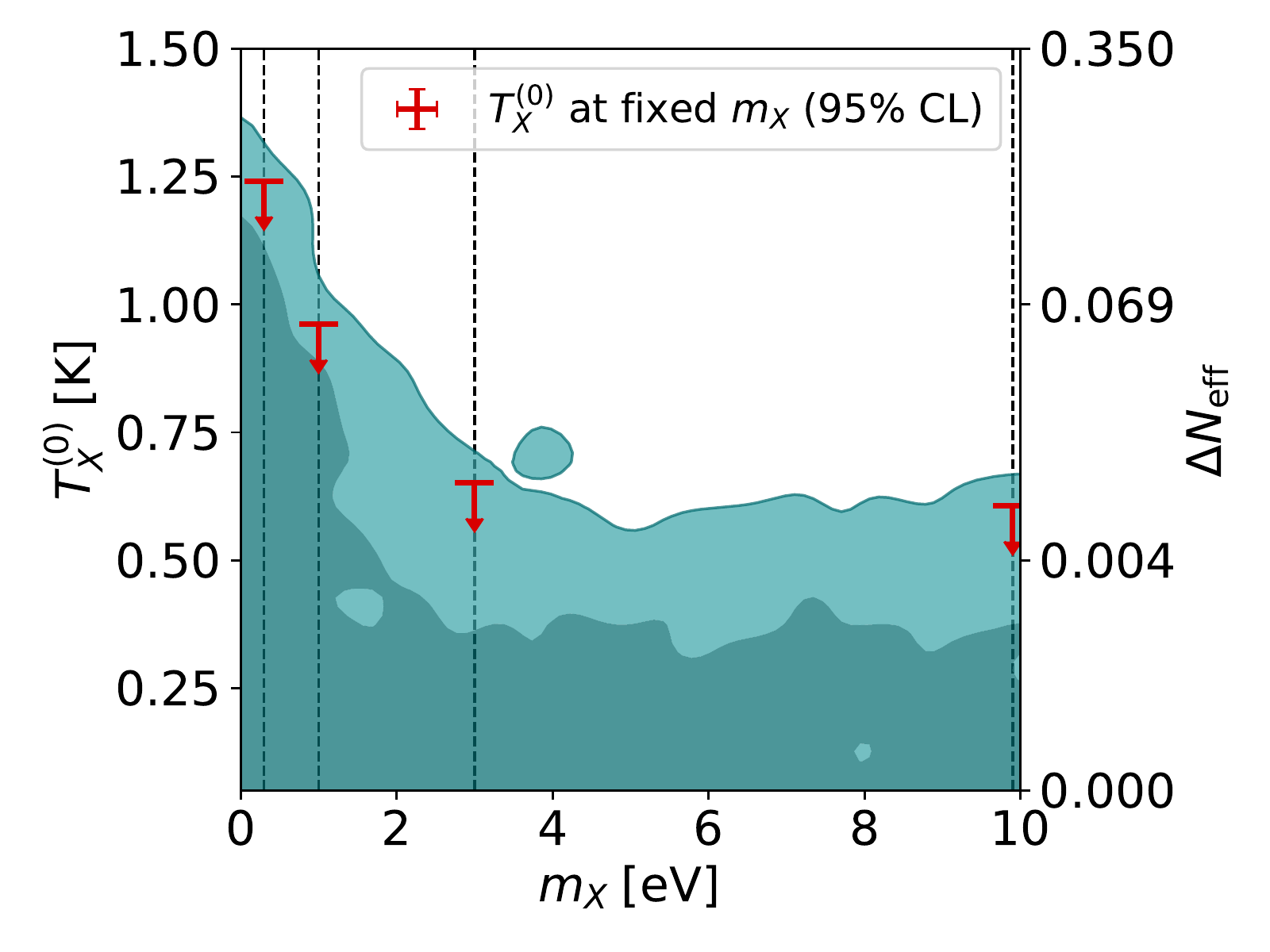}
    \caption{Constraints on the mass-temperature parameter space of an additional species of fermion light relic obtained via analysis of {\it Planck} 2018 TT+TE+EE+lensing, BOSS DR12 broadband spectra, and CFHTLens weak lensing data. Superimposed are the temperature constraints for a LiMR of fixed mass at 95\% CL. The joint analysis of CMB, LSS, and weak lensing data is critical to obtaining the most competitive constraints, as even in the limit of massless relic the shown constraints outperform $N_{\rm eff}$ constraints from CMB information alone -- and the incorporation of massive relic observables strengthens the constraining power considerably.}
    \label{fig:constraints}
\end{figure}

The present best constraints on the temperature-mass parameter space for massive light relics are shown in Figure~\ref{fig:constraints} (from~\cite{Xu:2021rwg}), obtained via a joint analysis of BOSS DR12 full-shape galaxy data, {\it Planck} 2018 temperature polarization and lensing anisotropies, and CFHTLens galaxy-galaxy ellipticity correlations. Presently, there is no evidence of light relics in our universe beyond the cosmic neutrinos predicted in the Standard Model. 
Focusing on the benchmark temperature of $T_X^{(0)}= 0.91$ K, expected of a minimally coupled thermal relic, the analysis of~\cite{Xu:2021rwg} constrains light relics scalars at masses of 11.2~\si{eV}, Weyl fermions at 2.26~\si{eV}, vectors at 1.56~\si{eV}, and Dirac fermions at 1.06~\si{eV}, all at 95\%~CL.

\subsection{Observational prospects}

Upcoming large-scale structure surveys such as the Dark Energy Spectroscopic Instrument (DESI), the {\it Vera Rubin Observatory} and Euclid will reduce the error bar on the sum of neutrino masses by a factor between $3$ and $5$~\cite{Sprenger:2018tdb,Mishra-Sharma:2018ykh}, depending on the uncertainties in the nonlinear modeling of galaxy and dark matter clustering. Next generation CMB experiments like the Simons Observatory~\cite{SimonsObservatory:2018koc} and CMB-S4~\cite{Abazajian:2019eic} will also contribute to further constrain neutrino masses~\cite{Brinckmann:2018owf,Dvorkin:2019jgs}. 
Given the similarity between neutrino masses and other LiMRs we can thus expect similar improvement for the latter, especially in the mass range at or below an \si{eV}, as indicated by recent forecasts~\cite{DePorzio:2020wcz}. 

Forthcoming surveys will be able to improve the bounds on LiMRs masses by another factor of two at least, thanks to the larger volume accessible.
Moreover, they will operate at redshift $z \gtrsim 2$, deep in the matter-dominated regime, and will therefore allow robust and strong constraints on the LiMR parameter space even in more extended cosmological models (for instance with evolving dark energy).

Within the next five years we could therefore be able to rule out the entire thermal gravitino parameter space~\cite{DePorzio:2020wcz, Xu:2021rwg} at the 2$\sigma$ level, and other motivated light relic scenarios such as Mirror Twin Higgs~\cite{Chacko:2018vss} (see Sec.~\ref{sec:TwinHiggs}), sterile neutrino (see Sec.~\ref{sec:SterileNeutrinos}), and hot axion models~\cite{Baumann:2016wac} will be likewise severely constrained within the next decade. 

\subsubsection*{Theoretical challenges}
For relics lighter than about an \si{eV} the modeling of the observables is well understood, and we thus expect current forecasts to be relatively accurate. 
The same cannot be said for all heavier LiMRs. As a relic becomes more massive, its nonlinear clustering becomes more important and it should be included in the prediction of the galaxy and weak lensing power spectra~\cite{LoVerde:2014rxa}. At the same time, heavier LiMRs will affect the nonlinear growth of standard cold dark matter (similarly to massive neutrinos~\cite{Ali-Haimoud:2012fzp,Senatore:2017hyk}). While both effects are qualitatively well understood, their technical implementation in perturbation theory has proven quite challenging, and we still lack exact and rigorous predictions of the cosmological observables. Developing analytical approaches to perturbation theory in presence of LiMRs will therefore be vital for a more robust detection or bound on the LiMRs parameter space.

\subsection{Relation to neutrino mass}
\label{sec:LiMR_Mnu}

For LiMRs with masses below about $\SI{0.2}{eV}$, the free-streaming length appears at large cosmological scales where LSS surveys sample few independent modes.  In this regime, the shape of the suppression of the matter power spectrum (see Figure~\ref{fig:pk_limr}) is difficult or impossible to resolve, implying that observations are only sensitive to the amplitude of the suppression.  These low-mass LiMRs therefore have observational signatures that can be well captured in terms of $\Delta \Neff$ and $\sum m_\nu$~\cite{Green:2021xzn}.  Similarly, models in which a fraction of the dark matter acquires a large velocity after recombination (e.g.~via DM-neutrino scattering~\cite{Green:2021gdc}) will lead to a suppression of clustering in a manner similar to that described above.  Despite the different physical mechanism, the observable consequences for cosmology are also similar to a change to $\sum m_\nu$~\cite{Green:2021xzn}.  Cosmological measurements of both $\Neff$ and $\sum m_\nu$ provide useful insights for dark sector physics.

\section{Dark sector complexity}

The Standard Model of particle physics is very far from minimal. The rich dynamics that characterize our visible universe arise due to several different matter states with different masses interacting under three different gauge forces, most notably for this discussion the confining strong force, which gives rise to hadronic as well as nuclear bound states, and the long-range electromagnetic force, which gives rise to atomic bound states, chemistry, and dissipation. The chiral nature of SM matter and the origin of fermion masses from spontaneous symmetry breaking can also be connected to the lightness of some of the leptons, i.e. the neutrinos. 
We probably would never have expected the existence of the SM if we did not live in it. What other strange sectors could be making up a similar $\mathcal{O}(10\%)$ fraction of the universe's matter budget? 
At its most basic level, this bottom-up plausibility argument sets the stage for considering \emph{Dark Complexity}, the possibility of hidden sectors with particles and forces that approach the SM in complexity, with possibly equally interesting dynamics in the early universe and today. 

The argument for Dark Complexity is not just based on plausibility, however. 
From a top-down point of view, many BSM theories postulate the existence of hidden sectors related by discrete symmetries to the SM~\cite{Chacko:2005pe,Barbieri:2005ri,Chacko:2005vw,Chacko:2016hvu,Craig:2016lyx,Chacko:2018vss,Chacko:2021vin,GarciaGarcia:2015pnn,Foot:2002iy,Foot:2003jt,Berezhiani:2003xm,Foot:1999hm,Foot:2000vy,Foot:2003eq, Foot:2004pa} (see also the dedicated Snowmass~2021 White Paper on early-universe model building~\cite{Snowmass2021:EarlyModels}). 
As will be discussed in more detail in~Section~\ref{s.naturalness}, this includes models of ``neutral naturalness'' such as the Mirror Twin Higgs that address the hierarchy problem of the Higgs boson mass~\cite{Chacko:2005pe,Barbieri:2005ri,Chacko:2005vw,Chacko:2016hvu,Craig:2016lyx,Chacko:2018vss,Chacko:2021vin,GarciaGarcia:2015pnn} by positing the existence of a hidden sector related to the SM by a $\mathbb{Z}_2$ or other discrete symmetry, resulting in analogues of electromagnetism, strong and weak forces as well as various matter states existing in the dark sector. Interestingly, while these dark sectors are qualitatively analogous to the SM in their ingredients, their dynamics can be very different owing to different effective values of the corresponding fundamental constants.
This can result in a fraction of dark matter being made up of dark baryons, with dark protons, dark electrons, dark photons, and even dark nuclear forces that are similar to their SM counterparts but with possibly different values for masses and interaction strengths. 
Another solution to the hierarchy problem that may give rise to similar phenomena is N-Naturalness~\cite{Arkani-Hamed:2016rle}. 

Regardless of any detailed UV-motivation, Dark Complexity is a broad umbrella term for a variety of DM models that have been considered in the literature, which feature significantly richer dynamics than the single collisionless WIMP or minimal axion. 
Light relics are often the result of dark complexity, precisely because there are many possible ways to realize very light or massless particles even just with analogues of dynamics that are contained in the SM. 

A simple example is composite DM~\cite{Cline:2021itd,SpierMoreiraAlves:2010err,Kribs:2016cew,Antipin:2015xia,Cline:2013zca,Khlopov:2008ty,Kribs:2009fy,Alves:2009nf,Ko:2017uyb}. In analogy to SM QCD, a composite DM sector could be as simple as some dark fermion(s) charged under a confining dark gauge force. All or part of DM could then be made up of a dark-hadronic bound state, with the details of its possible dynamics in our universe today ranging from being effectively collisionless to acting like self-interacting dark matter~\cite{Cline:2013zca,Ko:2017uyb,Ibe:2018juk} or even the nucleonic part of atomic dark matter (see below). 
However, it is noteworthy that the composite nature of the dark sector can itself give rise to a light degree of freedom if the confinement breaks a chiral symmetry, thereby ensuring the existence of a dark-pion pseudo- or exact Goldstone boson that can act like self-interacting dark radiation~\cite{Ko:2017uyb}. 

Another archetypical benchmark scenario of dark complexity is \emph{atomic dark matter}~\cite{Goldberg:1986nk,Kaplan:2009de, Kaplan:2011yj,Cline:2012is,Cline:2013pca,Fan:2013yva,Fan:2013tia,Fan:2013bea,Cyr-Racine:2013fsa,Rosenberg:2017qia,Ghalsasi:2017jna,Gresham:2018anj,Essig:2018pzq,Alvarez:2019nwt,Cline:2021itd,Cyr-Racine:2021alc,Blinov:2021mdk}, which 
postulates that dark matter contains at least two states with different masses and opposite charge under a \emph{dark electromagnetism} $U(1)_D$ gauge symmetry.
Straightforward generalizations include different charge ratios of the `dark nucleon' and the dark electron, and the possibility of many dark nucleon species. The latter scenario is especially motivated if atomic dark matter is realized together with composite DM, i.e. if the dark nucleons are actually dark-hadronic bound states which in turn form different nuclei, in analogy to the SM. 
This gives rise to one very obvious light relic: the dark photon, which in most atomic DM scenarios of interest is either very light or massless to allow for the formation of bound states. 
The dark photon contributes to the effective number of light species that can be measured from the CMB $\Delta \Neff \sim 4 (T_D/T_{CMB})^4$. For a dark sector that was in thermal equilibrium with the SM at the formation of the CMB, this is obviously excluded by {\it Planck} constraints~\cite{Planck:2018vyg}, but the strong temperature dependence means that even the normal entropy injections within the SM could supply sufficient dilution to make dark photons consistent with current data. A variety of other dilution mechanisms that lower the hidden sector temperature relative to the SM are also possible, see e.g.~\cite{Berezhiani:1995yi, Berezhiani:1995am, Adshead:2016xxj, Chacko:2016hvu, Craig:2016lyx}, making a dark photon contribution to $\Delta \Neff$ at the percent-level plausible and an attractive target for CMB-S4~\cite{CMB-S4:2016ple,Abazajian:2019eic}.
The long-range dark-electromagnetic interaction also causes dark acoustic oscillations  of the atomic DM component (see e.g.~\cite{Cyr-Racine:2013fsa, Chacko:2018vss}), which can leave oscillatory deviations in the matter power spectrum measurements compared to the $\Lambda$CDM expectation. 

Complete models can often feature several of the above mechanisms. For example, the Twin Higgs features a dark sector with composite twin protons (made of confining twin quarks) bound into twin atoms with twin electrons via twin electromagnetism. Depending on the nature of the discrete symmetry to the SM, it can also feature light twin neutrinos, as well as some source of dilution~\cite{Chacko:2016hvu, Craig:2016lyx} to satisfy CMB bounds on $\Delta \Neff$. Therefore, a variety of light states are present contributing to the above-discussed observables. It may well be that all these effects have to be present to resolve cosmological mysteries. For example, it has been shown that the $H_0$-$\sigma_8$ tension could be resolved by such a twin sector~\cite{Bansal:2021dfh,Cyr-Racine:2021alc,Blinov:2021mdk}, owing both to dark acoustic oscillation delaying structure formation at the required scale to ameliorate the $\sigma_8$ anomaly~\cite{Hildebrandt:2018yau}, and the interacting and non-interacting species of dark radiation contributing to $\Delta \Neff$, thereby alleviating the $H_0$ tension between the CMB measurement and the direct measurement today~\cite{Riess:2020fzl, Breuval:2020trd, Riess:2019cxk}.  

Models featuring dark photons and several matter states, like atomic or twin dark matter, are also interesting due to their rich dynamics in the universe today. 
They could easily be compatible with cosmological and self-interaction bounds, especially if the dark atoms make up a $\lesssim \mathcal{O}(10\%$) subcomponent of dark matter~\cite{Fan:2013yva,Cyr-Racine:2013fsa,Chacko:2018vss}. 
Like their SM counterparts, dark particles could cool and form structure, leading to the formation of a dark disk or dark microhalos~\cite{Fan:2013yva,Fan:2013tia,Ghalsasi:2017jna}.
On even smaller astrophysical scales, collapsing atomic dark matter would form mirror stars~\cite{Foot:1999hm, Foot:2000vy,Curtin:2019lhm, Curtin:2019ngc}, possibly emitting dark photons long past their Kelvin-Helmholtz time if dark nuclear interactions release energy in their cores. This opens a new window for astrophysical dark matter searches, since it has recently been shown~\cite{Curtin:2019lhm, Curtin:2019ngc} that mirror stars can produce electromagnetic signals that may be visible in optical/IR and X-ray telescopes if the dark photon has a tiny kinetic mixing with the SM photon. Such a small mixing $\epsilon F_{\mu \nu} F_D^{\mu \nu}$ is a renormalizable operator in the Lagrangian and generically expected to be produced at some loop level in the complete theory~\cite{Gherghetta:2019coi}, but even without such a coupling, purely gravitational signatures, like microlensing and gravitational waves, will be able to detect mirror stars in our galaxy and beyond~\cite{Winch:2020cju, Hippert:2021fch}.

Dark Complexity is difficult to study experimentally and theoretically due to the enormous range of possible dynamics it can generate in our universe, but any discovery would herald a new golden age of astrophysics and particle physics. A myriad of different signatures would need to be meticulously observed and understood to discern the nature of the dark sector. The robust nature of light dark relic signatures make them a crucial part of this puzzle. 

\subsection{Connections to Naturalness}
\label{s.naturalness}

It is well known that the Higgs boson in the Standard Model of particle physics introduces a naturalness puzzle in the presence of additional scales.  The fact that the weak scale is parametrically smaller than the scale of new physics (which could be as high as the Planck scale) cries out for an explanation.  There are two known theoretical mechanisms that can solve this problem for an arbitrarily high new physics scale: supersymmetry and Higgs compositeness.  Both approaches imply new physics states near the weak scale that should be accessible at collider experiments.  The current absence of evidence for new physics at the Large Hadron Collider (LHC) has motivated the exploration of new explanations for the lightness of the weak scale.

Our focus here is on the connection between light relics and electroweak naturalness.  To this end, we will focus our attention on two classes of models that both rely on the presence of hidden sector `copies' of the Standard Model.  (These copies could be exact duplicates or could be Standard Model-like `pseudo-copies.') For both classes of model, the copies could have associated light degrees of freedom that could impact cosmological observables.  These models provide examples where the first hint of the solution to the electroweak hierarchy problem could appear from cosmology.

\subsubsection{Twin Higgs}
\label{sec:TwinHiggs}

The Twin Higgs~\cite{Chacko:2005pe} is the simplest realization of the phenomenon of ``neutral naturalness'', in which partner particles appear near the weak scale consistent with natural expectations, but do not carry the same quantum numbers as their Standard Model counterparts~\cite{Chacko:2005pe,Barbieri:2005ri,Chacko:2005vw,Craig:2014aea}. These partner particles arise from discrete symmetries relating the Standard Model to additional sectors, as the discrete symmetry results in a larger accidental symmetry that protects the Higgs mass without requiring new states charged under the Standard Model. Although these models only stabilize the Higgs mass over one or two decades in energy (above which a more conventional solution such as supersymmetry or compositeness must appear), they suffice to reconcile the observed Higgs mass with the non-observation of new charged particles near the weak scale. While the new Standard Model-neutral partner particles are difficult to detect at the LHC, their cosmological signatures are among the most promising avenues for discovery. 

The discrete symmetry at play in the Twin Higgs is a $\mathbb{Z}_2$ symmetry relating the Standard Model to a mirror copy that contains the same field content and interactions. The existence of this mirror copy and a quartic coupling $\lambda$ between the Standard Model Higgs doublet and its mirror counterpart are sufficient to postpone the appearance of new charged particles associated with supersymmetry or compositeness by an amount $\lambda / \lambda_{\rm SM}$ compared to a theory without the $\mathbb{Z}_2$ symmetry. If the $\mathbb{Z}_2$ symmetry were exact, the scales $v$ and $f$ of Standard Model and mirror electroweak symmetry breaking would be identical. However, in the Twin Higgs model the couplings of the observed Higgs boson deviate from Standard Model predictions by an amount proportional to $v^2/f^2$, so $v \sim f$ is disfavored by LHC Higgs coupling measurements. Soft breaking of the $\mathbb{Z}_2$ allows $v \ll f$, but involves a fine-tuning proportional to $f^2/v^2$. This tuning is modest given current Higgs coupling measurements, but favors the Twin scale $f$ to lie as close as possible to the electroweak scale $v$.

The strongest constraint on the mirror Twin Higgs model comes from cosmology. The quartic coupling between the Standard Model Higgs doublet and its twin keep the two sectors in thermal equilibrium down to $\mathcal{O}(\si{GeV})$ temperatures~\cite{Chacko:2016hvu,Craig:2016lyx}. Assuming that the universe reheated above this temperature with otherwise typical cosmological evolution, the energy density in mirror photons and neutrinos corresponds to $\Delta N_{\rm eff} \approx 5.6$, such that the simplest mirror Twin Higgs is badly excluded by current bounds on $\Delta N_{\rm eff}$. There are then two broad possibilities consistent with current bounds: either the Twin sector is not an exact mirror copy (such that the Twin partners of the photon and neutrinos are heavy or decoupled~\cite{Craig:2015pha, Csaki:2017spo,Harigaya:2019shz}), or the energy densities in the Standard Model and Twin sector differ significantly in the early universe~\cite{Chacko:2016hvu,Craig:2016lyx, Curtin:2021alk}. While the former possibility can be arranged, it requires understanding why the $\mathbb{Z}_2$ symmetry is good for particles that couple strongly to the Higgs but bad for particles that do not. In contrast, the latter possibility preserves the $\mathbb{Z}_2$ symmetry in its entirety (modulo the soft breaking that leads to $v \ll f$) and predicts compelling cosmological signatures within reach of near-future experiments.

But how can the energy densities in the Standard Model and Twin sector differ significantly if the $\mathbb{Z}_2$ symmetry is only broken by the scale hierarchy $v \ll f$? A simple possibility is for a new particle coupling symmetrically to both sectors to decouple while relativistic, become non-relativistic and induce an era of matter domination, and then decay. Despite coupling symmetrically to both sectors in the underlying theory, its decay rate to each sector is inversely proportional to the corresponding scale. For example, if the new particle is a right-handed neutrino (motivated independently by neutrino mass generation), the ratio of its decay rates to the Standard Model and Twin sectors scales as~\cite{Chacko:2016hvu}
\begin{equation}
    \frac{\Gamma(N \rightarrow {\rm Twin})}{\Gamma(N \rightarrow {\rm SM})} \propto \frac{v^2}{f^2} \, . \label{eq:twinratio}
\end{equation}
As long as the new particle decays after the Standard Model and Twin sectors decouple, it will asymmetrically reheat the two sectors and suppress the energy density in the Twin sector proportional to the ratio in Eq.~\ref{eq:twinratio}. This is sufficient to reconcile the Twin Higgs with current limits on $\Delta N_{\rm eff}$ but provides a compelling target for future measurements: since the tuning of the weak scale is proportional to $f^2/v^2$ and the Twin contribution to $\Delta N_{\rm eff}$ is inversely proportional to the same ratio, improved sensitivity to $\Delta N_{\rm eff}$ probes the most natural parameter space of mirror Twin Higgs models. More broadly, the rich hidden sectors predicted by models of neutral naturalness provide a compelling target for cosmological probes of light relics connected to the stability of the weak scale.

\subsubsection[\texorpdfstring{$N$}{N}naturalness]{$N$naturalness}

\begin{figure}[t!]
    \centering
    \includegraphics[width=0.6\textwidth]{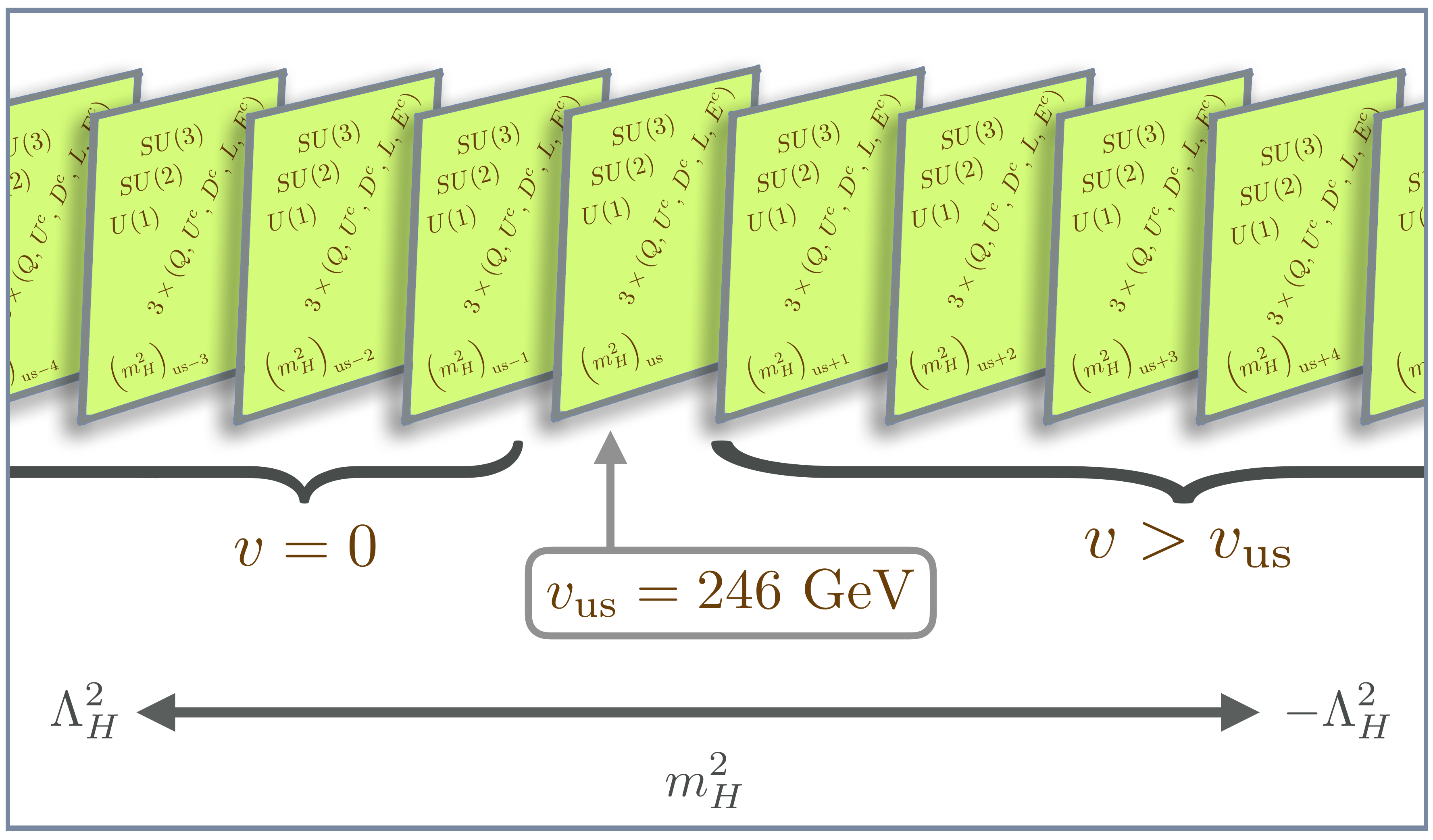}
    \caption{A sketch of the $N$naturalness setup.  There are $N$ copies of the Standard Model, where the only parameter that varies is the Higgs mass squared.  The accidentally lightest sector with a non-zero Higgs vev is our copy of the Standard Model.  Reproduced from Ref.~\cite{Arkani-Hamed:2016rle}.}
    \label{fig:nnaturalness}
\end{figure}

The starting point for the $N$naturalness model~\cite{Arkani-Hamed:2016rle} is to assume that there are $N$ copies of the Standard Model\footnote{It is not strictly required that the $N$ sectors be copies of the Standard Model.  This assumption is taken because it leads to a very predictive setup.}, which are identical except that their Higgs mass squared parameter take a random value between $-\Lambda_H^2$ and $\Lambda_H^2$, where $\Lambda_H$ is the effective UV cutoff that causes there to be fine-tuning in the Higgs mass parameter (see Figure~\ref{fig:nnaturalness}). This imples that we expect there to be one sector that accidentally has the smallest non-zero Higgs vev, which should parametrically be around the scale $\Lambda_H/\sqrt{N}$. The $N$naturalness approach is to find a dynamical reason that this sector is the one that dominates the energy density of the universe today.  This is the sense in which this model solves the hierarchy problem.  We live in the sector that is accidentally light, since this is the one that dominates the energy density of the universe.

To this end, we assume that when inflation ends, all of the energy density of the Universe is dominated by a particle that we call the ``reheaton''.  It is the decay of the reheaton that reheats the universe.  We assume that the reheaton is a gauge singlet, that it has a mass that is parametrically of order $\Lambda_H/\sqrt{N}$, and that its couplings are the most relevant possible that couple it to the Higgses of all the sectors.  A simple model with a scalar reheaton $\phi$ is therefore
\begin{equation}
    \mathcal{L} \supset - a \phi \sum |H_i|^2 - \frac12 m_\phi^2 \phi^2,
\end{equation}
where $H_i$ is the Higgs field for sector $i$.
For the lightest sector with a non-zero Higgs vev, the reheaton decays are unsuppressed.  For the rest of the sectors, the branching ratios for $\phi$ decays scale as $\Gamma \sim 1/m_h^2$ ($\Gamma \sim 1/m_{H}^4$), where the decays into sectors with non-zero (zero) vevs are dominated by decays to fermions (massless gauge bosons).  

\begin{figure}[t!]
    \centering
    \includegraphics[width=0.6\textwidth]{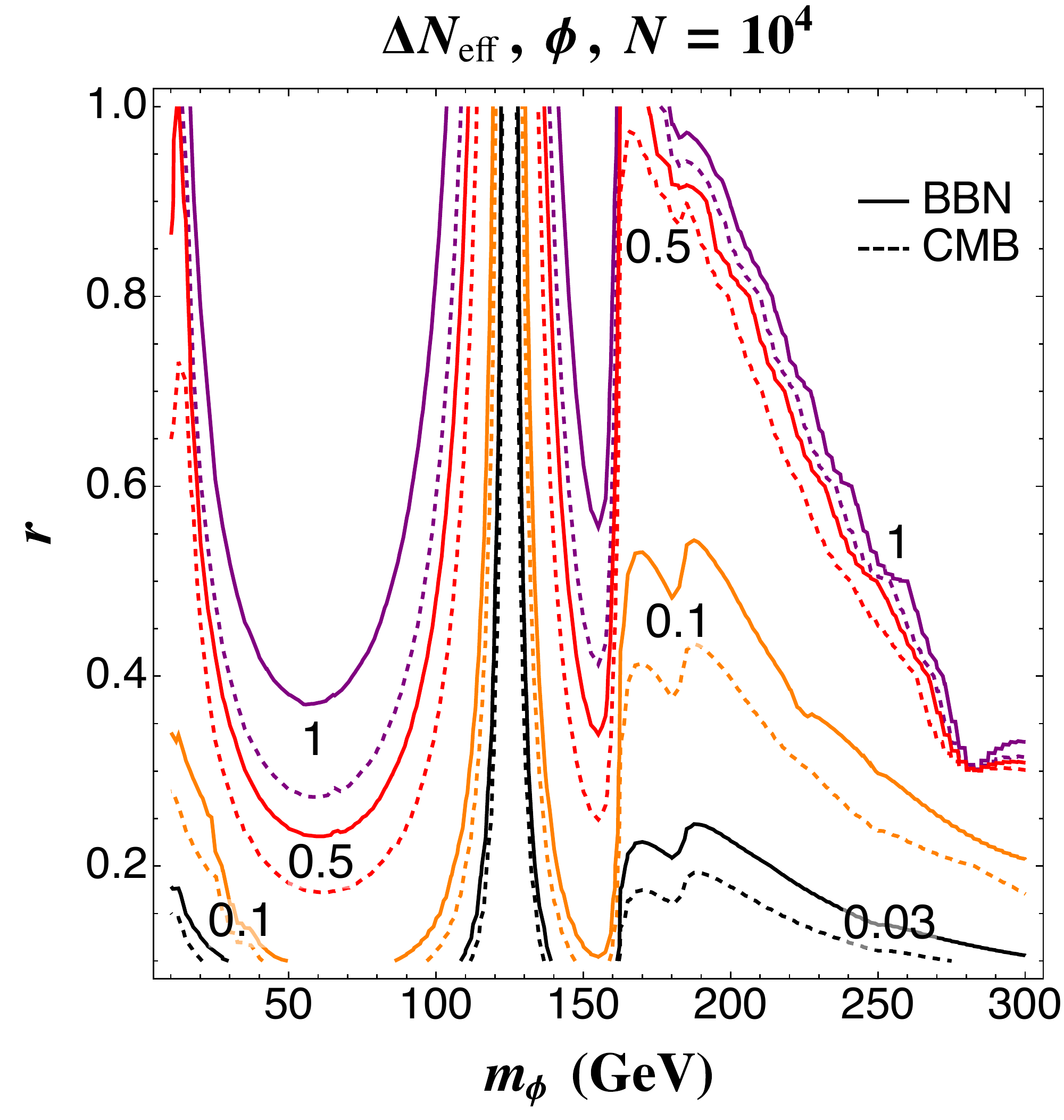}
    \caption{Contours of the contribution to $\Delta N_\text{eff}$ for BBN and the CMB in the $r$ versus $m_\phi$ (the mass of the reheaton) plane. Reproduced from~\cite{Arkani-Hamed:2016rle}.}
    \label{fig:nnaturalnessNeff}
\end{figure}

Although the decays into the other sectors are suppressed, they are non-zero, which implies that a variety of light relics will be produced.  In order to parametrize the observable impact this would have on $N_\text{eff}$, we introduce a parameter $r$:
\begin{equation}
    (m_H)^2_i = -\frac{\Lambda_H^2}{N}(2 i + r),
\end{equation}
where $i=0$ corresponds to ``us.''  This parameter $r$ tracks how close by the next sector is to us.  This parameter has a huge impact on the phenomenology, since it effectively controls how much energy density of the reheaton decays go into other sectors.  One can think of $r$ as a proxy for how much of an additional accident is required for the $N$naturalness model to be phenomenologically viable.  In Figure~\ref{fig:nnaturalnessNeff}, we show the predictions for $N_\text{eff}$ in the $r$ versus $m_\phi$ plane.  This shows that viable parameter space exists for the model. Additional constraints on the model (and an example model with a fermionic reheaton) are provided in~\cite{Arkani-Hamed:2016rle}.  

While these results are specific to the case of exact copies of the Standard Model, the generic expectation is that the hidden sectors would contain massless states.  Therefore, if the reheaton has a non-trivial branching ratio into the other sectors, there should be a non-trivial contribution to $N_\text{eff}$.
This is the sense in which probes of light relics correspond to testing the framework of $N$naturalness.

\subsection{Dark matter freeze-in } 

Dark matter (DM) could achieve its observed abundance through a mechanism known as freeze-in, where DM is produced non-thermally from SM particles in the thermal bath of the early universe annihilating and decaying. In particular, since the DM is non-thermal in this scenario, freeze-in is one of the few mechanisms for making sub-\si{MeV} DM from the SM plasma while still remaining consistent with bounds from BBN and $\Neff{}$ on thermal dark sectors~\cite{Boehm:2013jpa,Nollett:2013pwa,Green:2017ybv,Knapen:2017xzo,Sabti:2019mhn,Giovanetti:2021izc,An:2022sva}.

A classic example of a DM candidate born via freeze-in is the Dodelson-Widrow sterile neutrino~\cite{Dodelson:1993je}, but freeze-in is now known to be a more general mechanism that can be realized in a number of different models in the limit of small DM-SM couplings~\cite{Asaka:2005cn,Asaka:2006fs,Gopalakrishna:2006kr,Page:2007sh,Hall:2009bx,Bernal:2017kxu}. Producing the observed DM relic abundance without thermalizing the DM itself implies a tiny value for the coupling constants of order $\sim10^{-12}$~\cite{Chu:2011be, Dvorkin:2019zdi}, which is difficult to target with accelerator searches. However, if the DM-SM coupling is via a light mediator, this aids in the experimental observability of this candidate in direct detection experiments, since scattering will scale like $v^{-4}$ for velocity $v$, which is $v\sim 10^{-3}c$ at the Earth's location in the Milky Way (MW). For this reason, DM freeze-in via a light mediator is one of the key benchmarks for direct searches for DM where the same couplings determine the relic abundance and laboratory signals~\cite{Essig:2011nj,Essig:2012yx,Essig:2015cda,Aguilar-Arevalo:2019wdi,Barak:2020fql,Amaral:2020ryn,Agnes:2018oej,Hochberg:2017wce,Geilhufe:2019ndy,Coskuner:2019odd,Knapen:2017ekk,Griffin:2018bjn,Griffin:2019mvc,Berlin:2019uco}. Because of the temperature scaling of freeze-in production of DM via a light mediator, most of the DM is produced at late times, making this scenario relatively insensitive to initial conditions (unless the mediator is thermally populated at early times, see e.g. Ref.~\cite{Fernandez:2021iti}). 

If the mediator responsible for freeze-in is able to come into thermal equilibrium with the Standard Model plasma, it necessarily contributes $\Delta \Neff \geq 0.09$~\cite{Green:2017ybv,Knapen:2017xzo}. A cosmological exclusion of~$\Delta\Neff$ at this level would dramatically improve on direct constraints on DM-baryon interactions from cosmology and astrophysics (or meson decay searches) by many orders of magnitude~\cite{Green:2017ybv}. In this sense, the constraints on mediators imposed by measurements of $\Neff$ is one way in which the light relic density provides useful insights into the physics of the dark sector. On the other hand, if the light mediator is the SM photon, then $\Neff$ remains unchanged and if the mediator is a kinetically mixed dark photon then it will generically not thermalize with the SM due to a strong in-medium suppression of producing ultralight dark photons in a plasma~\cite{An:2013yfc}. 
Therefore, in the most viable models of freeze-in via an ultralight mediator, the DM is effectively millicharged; freeze-in is one of the few ways of making millicharged DM, since the couplings for standard thermal freeze-out are excluded by other constraints on millicharged DM~\cite{McDermott:2010pa}. 

Freeze-in via a light mediator therefore sits at the nexus of many interesting DM properties: it can make millicharged DM at sub-\si{MeV} mass scales without being excluded by present constraints, in a way that is insensitive to initial conditions and that can be imminently tested by a host of proposed DM direct detection experiments. It is thus timely to consider how the cosmological effects of DM freeze-in provide an observational handle that is complementary to other probes. There are two main effects that lead to observable departures from standard $\Lambda$CDM: (1) the portal responsible for creating the DM necessarily implies a drag force between the DM and the photon-baryon fluid before and during recombination, altering anisotropies seen in the CMB and (2) the DM is born with a nonthermal, high-velocity phase-space distribution, which suppresses clustering on small scales. For both observable effects, the full non-thermal velocity distribution of the DM is of critical relevance. The DM-SM scattering cross section responsible for the drag effect scales like $v^{-4}$ and depends strongly on the low-velocity part of the non-thermal distribution, while the suppressed growth of structure is sensitive to DM in the high-velocity tail of the non-thermal distribution.

For sub-\si{MeV} DM, the main channels for freeze-in production as computed in Ref.~\cite{Dvorkin:2019zdi} are from annihilation of electrons $e^+ e^- \to \chi \bar \chi$ and from plasmon decays $\gamma^* \to \chi \bar \chi$ (the decay of photons that acquire an effective in-medium plasma mass, see e.g.~\cite{Braaten:1993jw}). The initial non-thermal phase space can potentially become thermal at late times through DM-DM interactions; there is a window for this to occur without violating bounds on self-interacting DM~\cite{Tulin:2017ara}. 

If the DM does not self-interact substantially and the primordial phase space is preserved, then the implication for effect (1) is that constraints on millicharged DM assuming cold initial conditions (e.g.~\cite{Dvorkin:2013cea,Boddy:2018wzy,Slatyer:2018aqg}) will be relaxed somewhat because DM that is born via freeze-in through a light mediator is born relativistic; this is even more true for DM that does thermalize with itself through self-interactions. Current {\it Planck}~2018~\cite{Aghanim:2019ame} constraints on DM freeze-in through a light mediator are 18.5 and 19.3~\si{keV} for these respective scenarios; in the future, CMB-S4~\cite{CMB-S4:2016ple,Abazajian:2019eic} could probe up to masses of 28.8 and 28.1~\si{keV}, respectively~\cite{Dvorkin:2020xga}.

\begin{figure}
    \centering
    \includegraphics[width=0.7\textwidth]{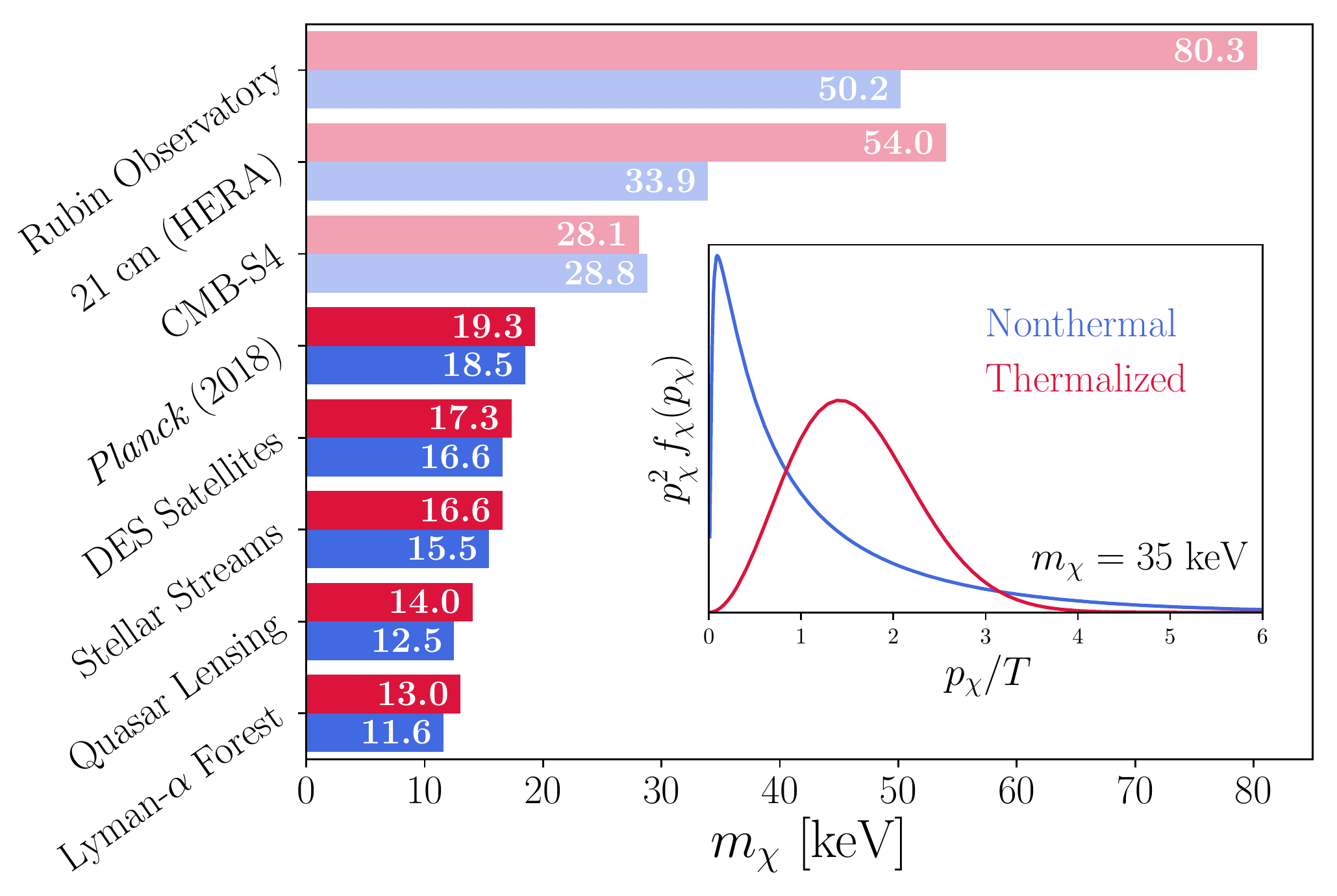}
    \caption{Cosmological 95\% bounds on DM that is produced by freeze-in through a light mediator. Dark shaded bars correspond to excluded DM masses while light ones correspond to projected future reach. These differ depending on whether the DM phase space remains nonthermal (blue) or thermalizes through self-scattering (red). Inset: the primordial DM phase space from the freeze-in mechanism (blue) and the resulting phase space from thermalization (red). Figure reproduced from Ref.~\cite{Dvorkin:2020xga}.}
    \label{fig:freezein_dm}
\end{figure}

Meanwhile, for effect (2), if the DM retains its primordial freeze-in phase space distribution and does not interact non-gravitationally, then it effectively behaves as a massive neutrino species with a unique velocity distribution. On the other hand, if freeze-in DM thermalizes at a later time, it behaves as a collisional fluid. In either case, advances in constraining warm DM can be used concurrently for the purposes of constraining freeze-in via a light mediator. For example, limits on warm DM have been set using the Lyman-$\alpha$ forest~\cite{Irsic:2017ixq,Yeche:2017upn,Baur:2017stq} and inferences about the subhalo mass function from quasar strong lensing~\cite{Hsueh:2019ynk,Gilman:2019nap}, stellar streams~\cite{Banik:2019cza,Banik:2019smi}, and MW satellite galaxies as observed by the Dark Energy Survey (DES)~\cite{Nadler:2019zrb,Nadler:2020prv}. The strongest constraints on freeze-in via a light mediator come from DES, corresponding to 16.6 and 17.3~\si{keV} for nonthermal and thermalized freeze-in distributions~\cite{Dvorkin:2020xga} (see Refs.~\cite{Ballesteros:2020adh,DEramo:2020gpr,Baumholzer:2020hvx,Dienes:2021cxp,Decant:2021mhj} for complementary treatments of \si{keV}-scale non-thermal DM and the effect on structure formation). Note that analogous analyses with DES~\cite{Nadler:2020prv} have placed a lower limit on the mass of Dodelson-Widrow sterile neutrinos of 50~\si{keV} and have similarly excluded nearly all the allowed parameter space for the Shi-Fuller mechanism for the production of sterile neutrinos~\cite{Shi:1998km}. In the near future, advances in analyzing the Lyman-$\alpha$ forest~\cite{Rogers:2020cup,Pedersen:2020kaw} as well as joint analyses of different probes~\cite{Enzi:2020ieg,Nadler:2021dft} may strengthen these bounds up to the $\sim 30$~\si{keV} level~\cite{Dvorkin:2020xga}. On longer timescales, these bounds on freeze-in can extend up to $50 - 80$~\si{keV} using inferences about small-scale halos from the Rubin Observatory~\cite{Drlica-Wagner:2019xan} and Hydrogen Epoch of Reionization Array~\cite{Munoz:2019hjh} (see Figure \ref{fig:freezein_dm}).

\section{Sterile neutrinos}
\label{sec:SterileNeutrinos}

\subsection{Status and motivation for a light sterile neutrino}

\begin{figure}[thbp]
   \centering
   \includegraphics[width = 0.408\textwidth]{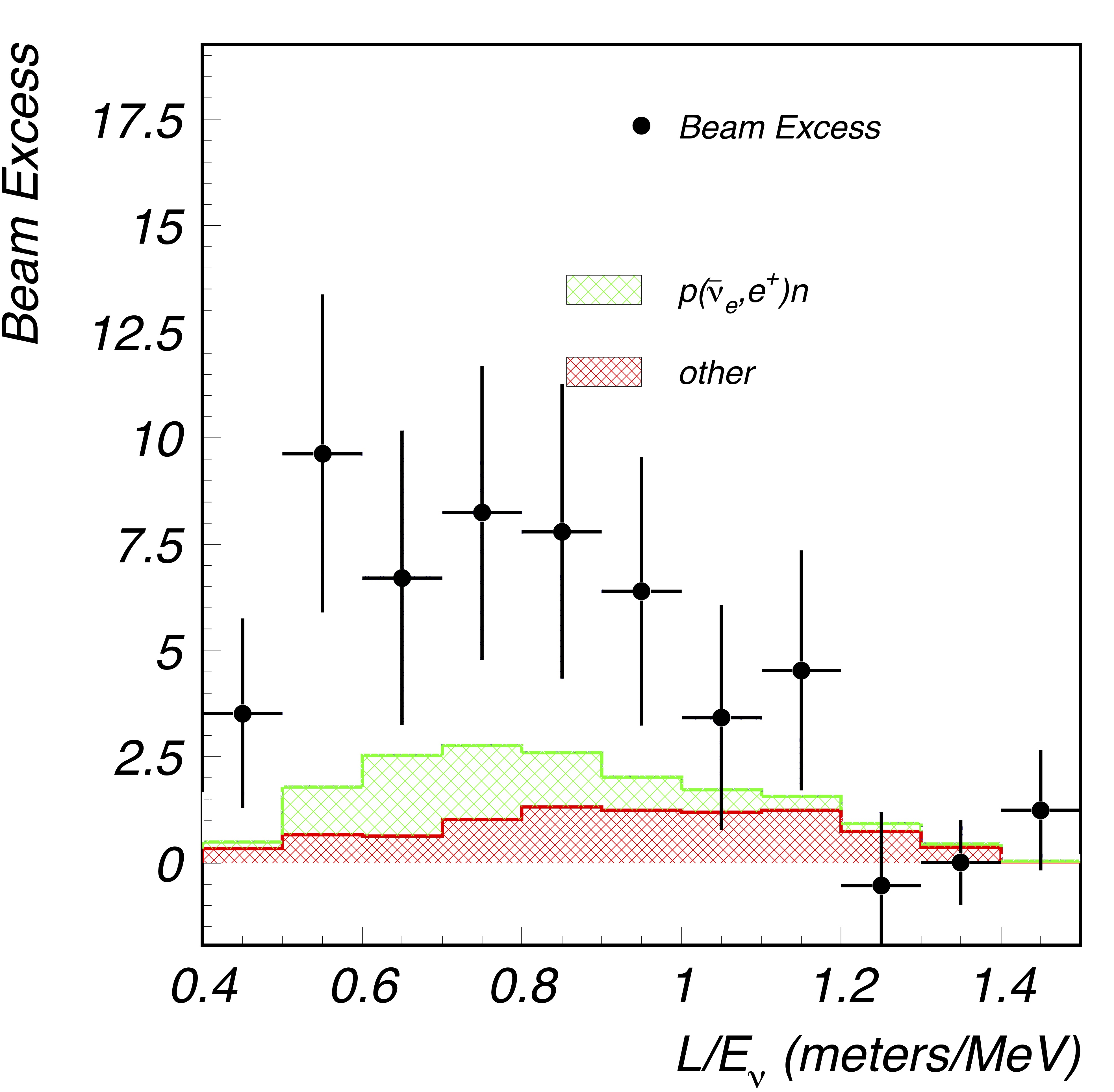}
   \includegraphics[width = 0.58\textwidth]{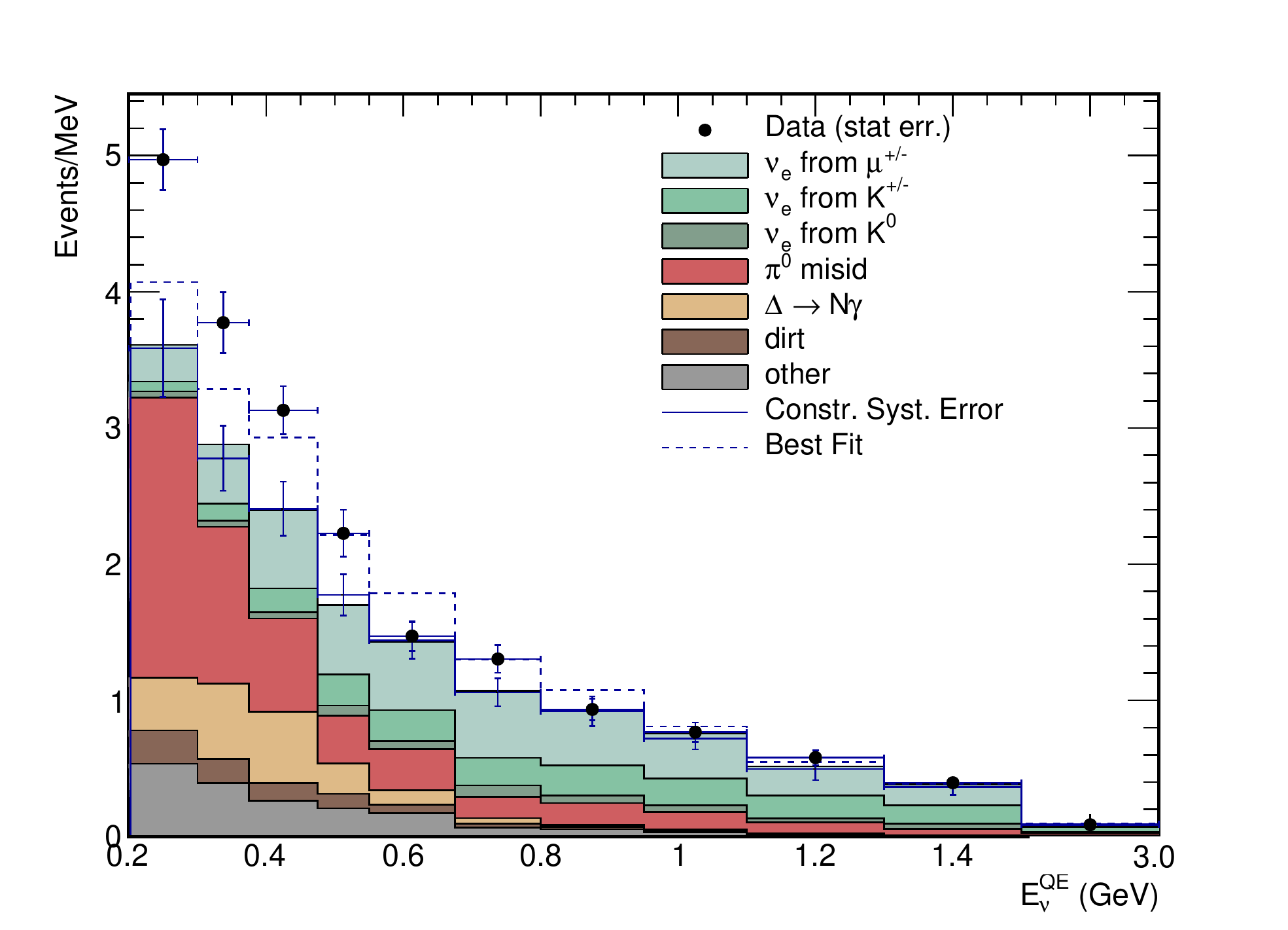}
   \caption{\textbf{\textit{The two elephants in the room: the LSND and MiniBooNE anomalous excesses.}}
   Left figure shows the LSND excess~\cite{LSND:1996ubh,LSND:2001aii}, while right figure shows the MiniBooNE excess~\cite{MiniBooNE:2013uba,MiniBooNE:2018esg}.
   }
   \label{fig:exp_excess}
\end{figure}

The LSND short-baseline neutrino experiment operated in the 90s observed an excess of electron-antineutrino events using a pion decay-at-rest beam~\cite{LSND:1996ubh,LSND:2001aii}; see Figure \ref{fig:exp_excess} (left).
Given the energies and baselines relevant for LSND and known neutrino oscillation lengths, this observation cannot be accommodated in the Standard Model with three massive neutrinos.
The simplest explanation to this observation is to introduce a new, heavier neutrino mass state of order $\SI{1}{eV}$~\cite{Abazajian:2012ys}.
This new mass state is predominantly of sterile flavor --- \textit{i.e.} the new flavor state does not participate in the Standard Model interactions --- to avoid stringent constraints from the invisible $Z$ decay~\cite{ALEPH:2005ab}.

\begin{figure}[thbp]
   \centering
   \includegraphics[width = 0.85\textwidth]{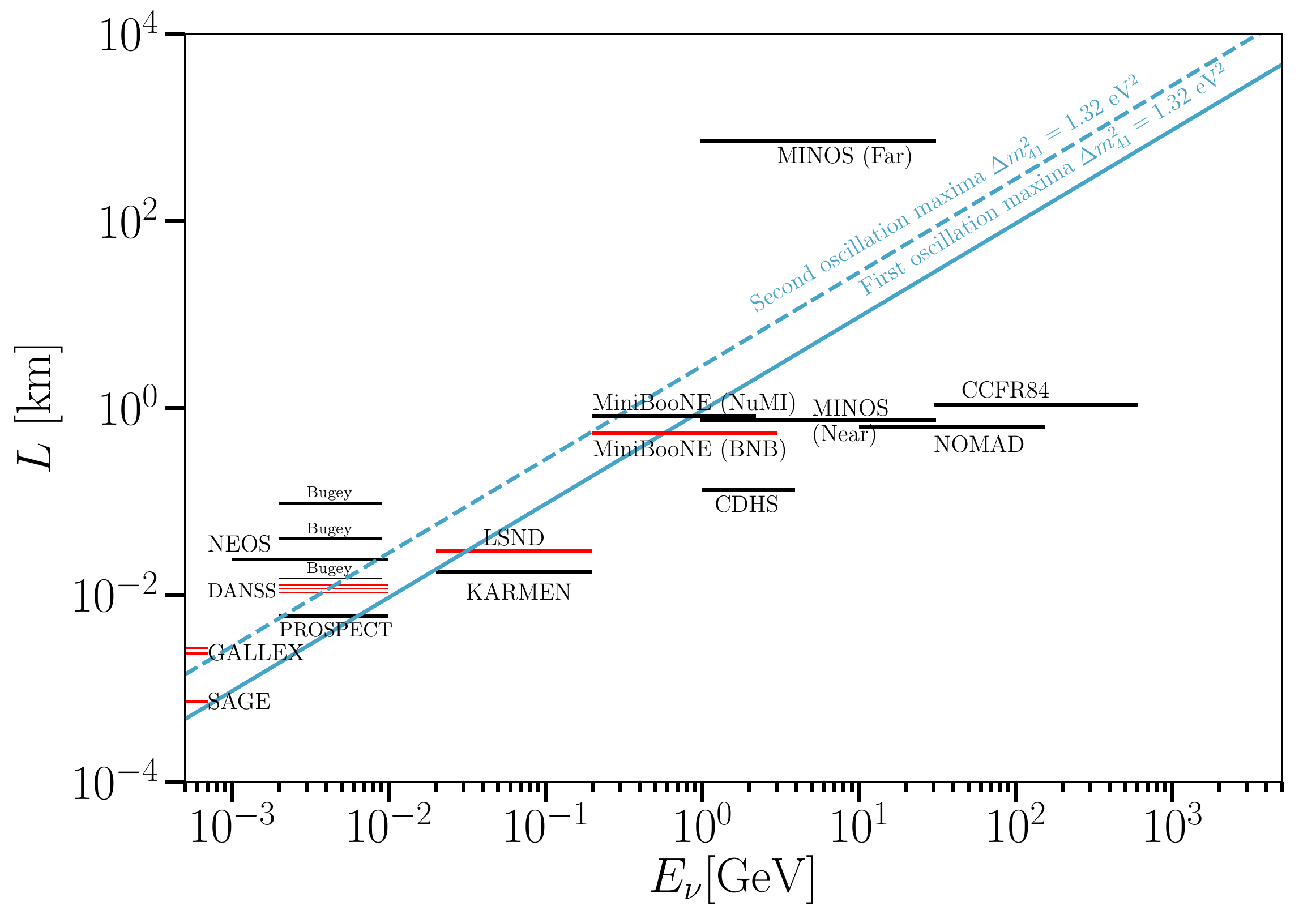} 
   \caption{\textbf{\textit{Neutrino experiments that are relevant for the short-baseline anomalies.}}
   Missing from this figure are the atmospheric neutrino experiments (SuperKamiokande, IceCube, and ANTARES), which populate the top-right corner and the solar neutrino experiments, which probe the top-left corner of this phase space.
   Experiments that have seen anomalous signals are shown in red. Figure reproduced from Ref.~\cite{Diaz:2019fwt}.
   }
   \label{fig:experiments_overview}
\end{figure}

This anomalous electron-neutrino appearance has motivated experiments to confirm the light sterile neutrino hypothesis.
These experiments range in energies from \si{MeV} to \si{TeV} and have baselines from meters to the diameter of the Earth as shown in Figure~\ref{fig:experiments_overview}.
At the lowest energies they search for electron-neutrino disappearance using reactor neutrinos~\cite{Mention:2011rk,Giunti:2010zu,Bahcall:1994bq,Bahcall1997,SAGE:1998fvr,Abdurashitov:2005tb,Barinov:2021asz,DayaBay:2016qvc, DayaBay:2016ggj,NEOS:2016wee,Stereo66,solid,Serebrov:2020kmd,RENO:2020uip, PROSPECT:2020sxr,Barinov:2021mjj,DANSS:2018fnn,Danilov:2021oop}, at the intermediate energy ranges $\mathcal{O}(\SI{1}{GeV})$ they use pion decay-in-flight beams~\cite{Armbruster:2002mp,Cheng:2012yy,MiniBooNE:2013uba,MiniBooNE:2018esg,MicroBooNE:2021jwr,MicroBooNE:2021nxr,MicroBooNE:2021rmx,MicroBooNE:2021sne,MINOS:2017cae}, while at the highest energies they use neutrinos produced in the atmosphere from pion and kaon decay~\cite{Aartsen:2017bap,IceCube:2016rnb,IceCube:2020tka,IceCube:2020phf,Abe:2014gda}.

Notably, the MiniBooNE experiment~\cite{MiniBooNE:2013uba,MiniBooNE:2018esg}, operating at Fermilab, searched for electron-neutrino and -antineutrino appearance at a ratio of baseline to energy similar to LSND.
MiniBooNE has reported an excess of electron-like events that is consistent with the observation of LSND and results in a combined significance that is above $5\sigma$~\cite{MiniBooNE:2018esg}; see Ref.~\ref{fig:exp_excess}.
Recently, the MicroBooNE experiment operating on the same beam as MiniBooNE has performed searches for an excess of electrons using a data-driven model that predicts the expected number of events in MicroBooNE given the excess observed in MiniBooNE.
MicroBooNE performed analyses~\cite{MicroBooNE:2021jwr,MicroBooNE:2021nxr,MicroBooNE:2021rmx,MicroBooNE:2021sne} searching for different final states and did not find significant evidence of electron-neutrino appearance given the shape predicted by the data-driven model.
However, when these results are interpreted in the context of light sterile neutrinos, they result in either weak hints of electron-neutrino disappearance~\cite{Denton:2021czb} or weak constraints~\cite{Arguelles:2021meu,MiniBooNE:2022emn}.

Additionally, measurements of electron-antineutrino disappearance in reactor experiments using ratios of events measured at different positions have shown hints of oscillations~\cite{Serebrov:2020kmd} compatible with light sterile neutrinos.
Recently, the BEST experiment operating in Russia and using a radioactive source has reported electron-antineutrino disappearance in the region of interest at a significance that also exceeds $5\sigma$~\cite{Barinov:2021asz}; however, at a mixing angle that is in tension with other reactor measurements~\cite{Berryman:2021yan}.
Unfortunately, searches for muon-neutrino disappearance have failed to find significant evidence for a light sterile neutrino.
Notably the MINOS+~\cite{MINOS:2017cae} and IceCube neutrino experiments~\cite{IceCube:2016rnb,IceCube:2020tka,IceCube:2020phf} have place stringent constraints on the muon-neutrino component of the heavy mass state.

\begin{figure}
    \centering
    \includegraphics[width=0.3\textwidth]{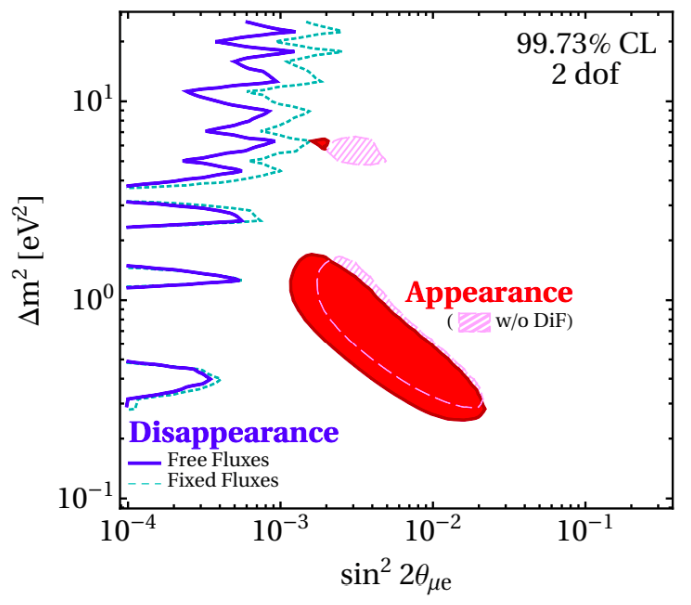}~ 
    \includegraphics[width=0.335\textwidth]{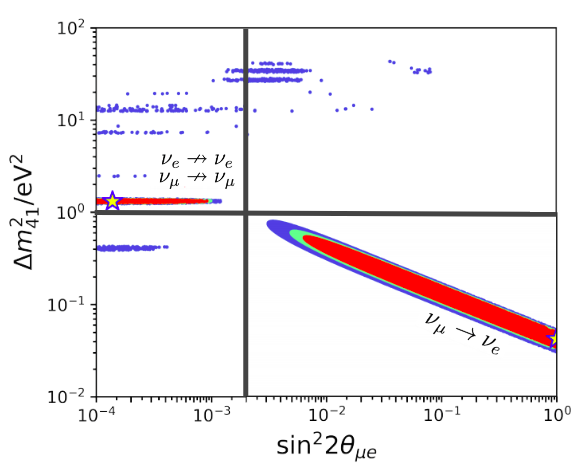}~
    \includegraphics[width=0.3\textwidth]{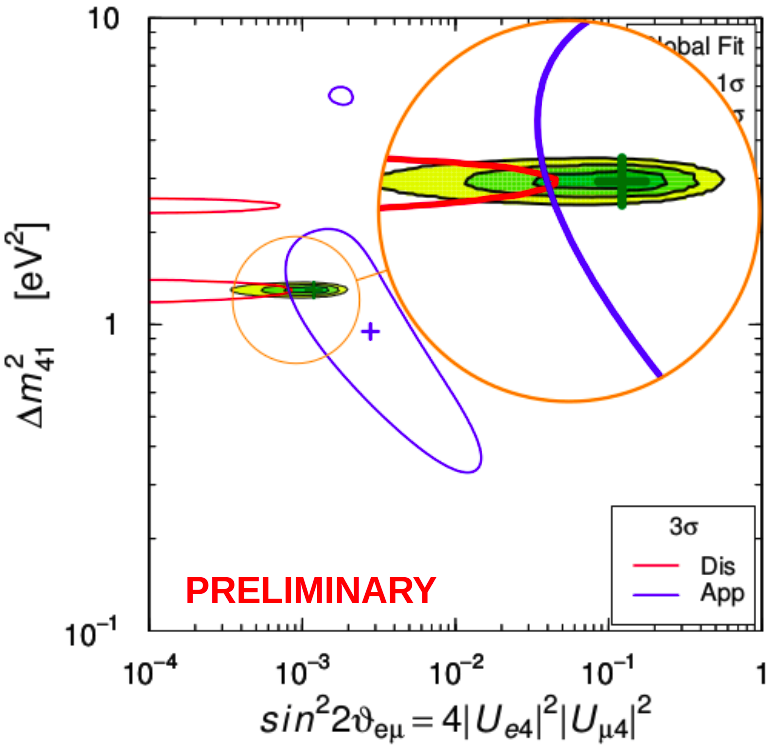}
    \caption{\textbf{\textit{Tension between the appearance and disappearance data sets.}}
    The $\nu_{\mu}\to\nu_e$ appearance amplitude in a 3+1 model is shown in the horizontal axis. 
    Allowed or preferred regions for appearance and disappearance global combinations from three different global-fit groups are shown in left~\cite{Dentler:2018sju}, center~\cite{Diaz:2019fwt}, and right~\cite{ternes_christoph_andreas_2018_2642337,Giunti:2019aiy} plots.
    See the references for the details.
    }
    \label{fig:tension}
\end{figure}

The situation described above has been studied in global analyses to the relevant neutrino data~\cite{Dentler:2018sju,Giunti:2019aiy,Diaz:2019fwt,Boser:2019rta} and have found that the regions preferred by the appearance experiments are in severe tension with those allowed by disappearance experiments; see Figure~\ref{fig:tension}.
This implies that even though the data significantly prefers something beyond the known three neutrinos, the light sterile neutrino hypothesis is inconsistent with the global data and is disfavored as a solution to the short-baseline anomalies.
This situation has propelled the community to explore scenarios beyond the light sterile neutrino.
These include introducing additional new physics that aims to resolve the tension between appearance and disappearance such as neutrino decay~\cite{PalomaresRuiz:2005vf,Moss:2017pur,Dentler:2019dhz,deGouvea:2019qre}, additional neutrino mass states~\cite{Diaz:2019fwt}, and secret neutrino interactions with non-minimal heavy neutral leptons~\cite{Gninenko:2009ks,Gninenko:2010pr,Masip:2012ke,Radionov:2013mca,Blennow:2016jkn,Ballett:2018ynz,Bertuzzo:2018itn,Arguelles:2018mtc,Liao:2018mbg,Denton:2018dqq,Ballett:2019pyw, Fischer:2019fbw, Jones:2019tow,Vergani:2021tgc}, among others~\cite{Strumia:2002fw, Murayama:2000hm, Barenboim:2002ah, GonzalezGarcia:2003jq, Barger:2003xm, Diaz:2010ft, Barenboim:2004wu,Sorel:2003hf, Farzan:2008zv,deGouvea:2006qd,Pas:2005rb, Carena:2017qhd, Hollenberg:2009ws,Zurek:2004vd, Kaplan:2004dq, Schwetz:2007cd, Armbruster:2003pq, Gaponenko:2004mi,Bai:2015ztj,Hollenberg:2009tr,Li:2007kj,Nelson:2010hz,Liao:2016reh, Akhmedov:2010vy, Papoulias:2016edm}.

At present, this confusing experimental situation requires further measurements to be resolved and differentiate between the vanilla light sterile neutrino solution and new proposed solutions to the short-baseline puzzle.

\subsection{Cosmological light sterile neutrinos}

Standard cosmological scenarios predict that the three active neutrino states remain in thermodynamic equilibrium with other Standard Model particles in the early universe through weak interactions.
As the universe expands eventually the neutrino interaction rate falls below the Hubble expansion decoupling the neutrinos from the plasma. 
At the time of neutrino decoupling, the neutrino temperature $\mathcal{O}(\SI{1}{MeV})$ is significantly larger than the known neutrino masses, $<\SI{1}{eV}$, and these can be considered a relativistic ensemble.
The number of effective relativistic species at these time, $\Neff$, is expected to be close to three with small corrections~\cite{Froustey:2020mcq,Bennett:2020zkv}.
Additionally, relativistic degrees of freedom, such as additional neutrino states, would contribute to $\Neff$.

\begin{figure}[t]
    \centering
      \includegraphics[width=0.45\textwidth]{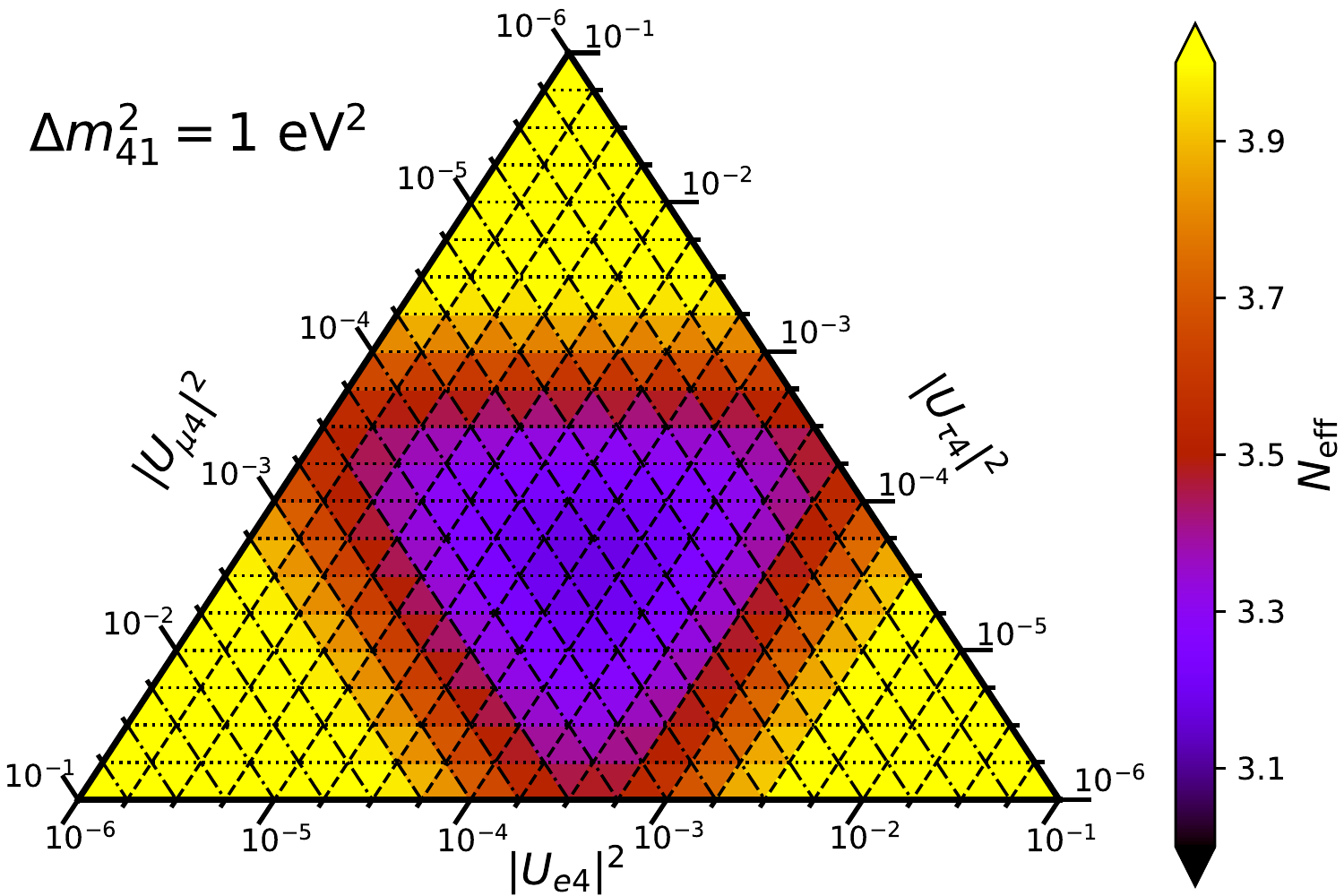}
      \includegraphics[width=0.49\textwidth]{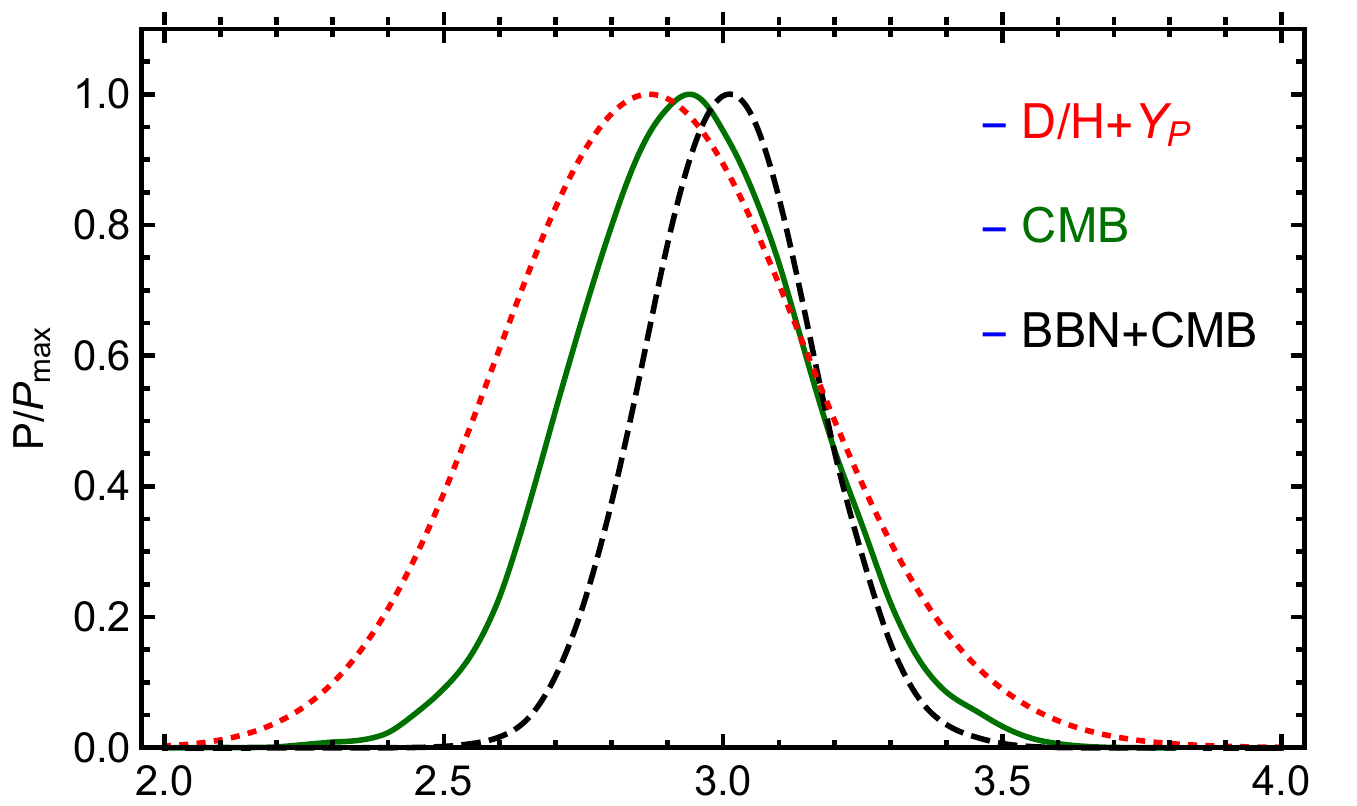}
    \caption{\textbf{\textit{Expected value of $N_\mathrm{eff}$ and current constraints.}} The left panel shows the expected value of $\Neff$ for a light sterile neutrino with a mass-squared-difference compatible with the observed anomalies~\cite{Gariazzo:2019gyi}. The right panel shows recent current constraints on $\Neff$~\cite{Pitrou:2018cgg}.}
    \label{fig:n_effective}
\end{figure}

\begin{figure}[t]
    \centering
      \includegraphics[angle=0,width=0.49\textwidth]{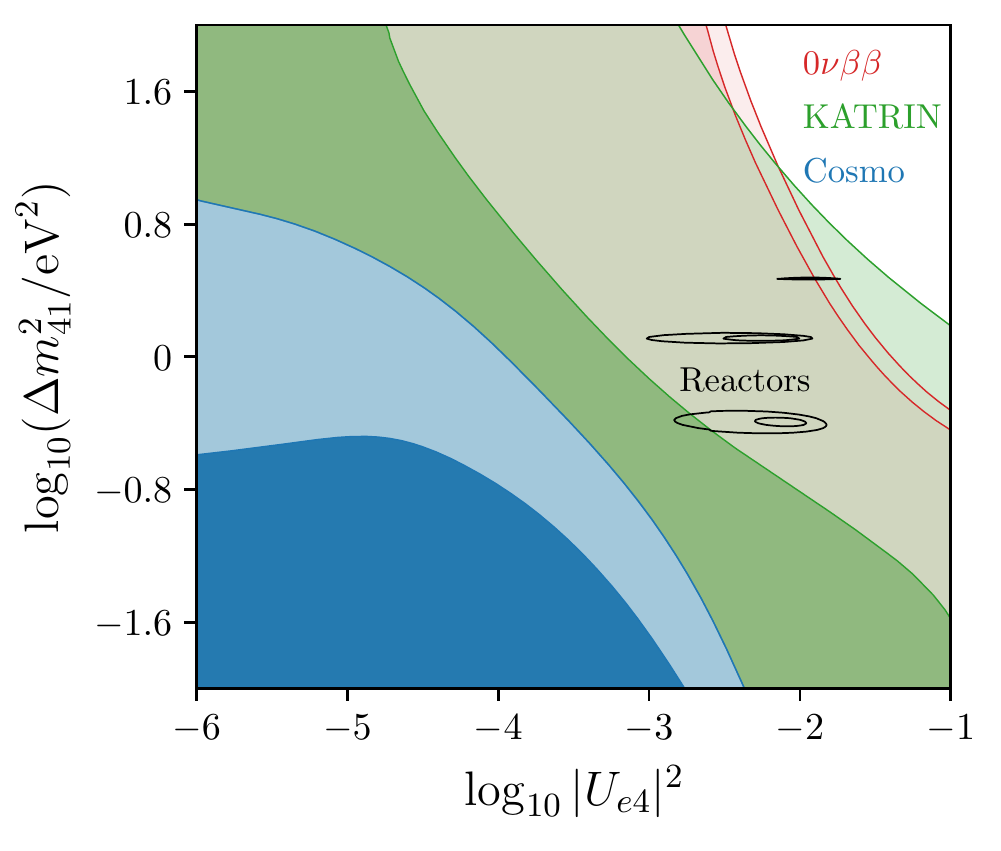}
      \includegraphics[angle=0,width=0.49\textwidth]{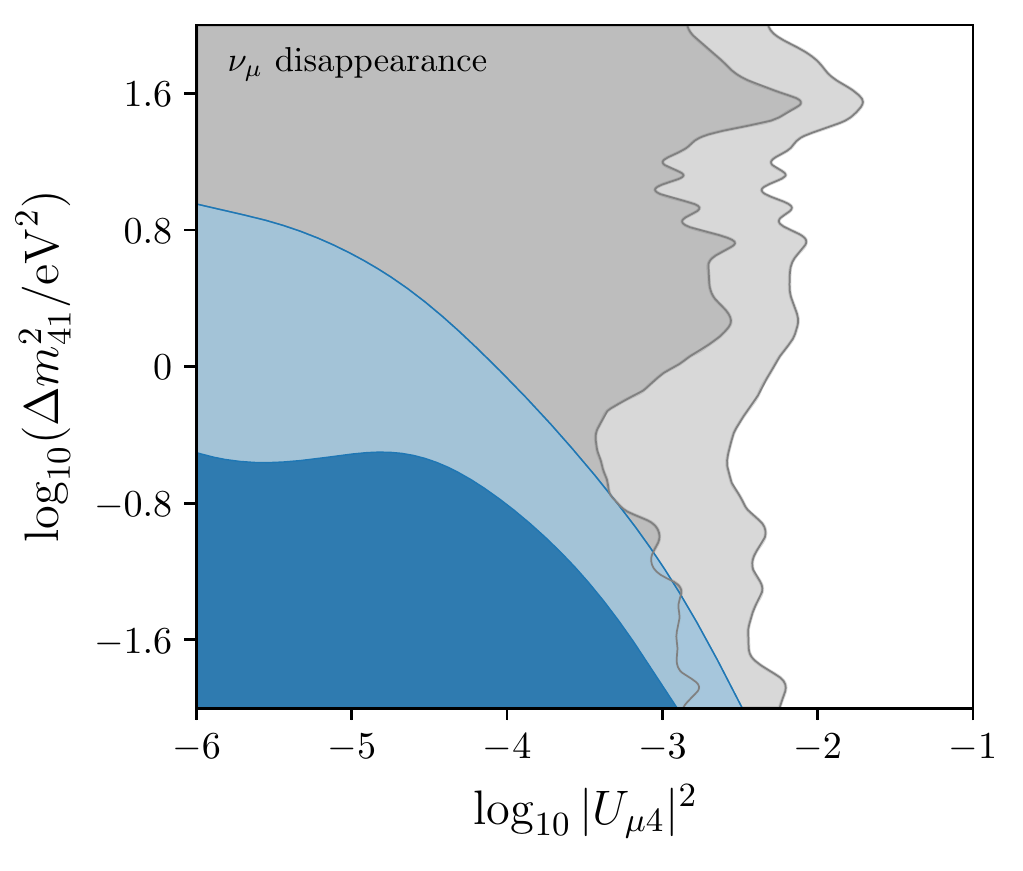}
    \caption{\textbf{\textit{Terrestrial and cosmological constraints on light sterile neutrinos.}} Left (right) shows mass-squared differences versus the electron (muon) mixing. Ranges allowed by cosmology are shown in blue, measurements from direct neutrino mass measurements in green, and constraints from oscillations in gray. Preferred regions are indicated as small circles. Figure reproduced from~\cite{Hagstotz:2020ukm}.}
    \label{fig:cosmo_constraints}
\end{figure}

Light sterile neutrinos motivated by the anomalies previously discussed are expected to thermalize in the early universe and contribute to $\Neff$.
If the mixing between active and sterile neutrinos is large enough, these will be produced in the early universe via a combination of active-sterile neutrino oscillations and collisions of Standard Model particles.
Concretely, active neutrinos begin to oscillate when the oscillation frequency, $\Delta m^2/2E$, is larger than the Hubble expansion rate, $H(t)$, transforming into sterile neutrinos.
Collisions can also produce the heavier mass state since it has a small active flavor mixing, increasing the population of light sterile neutrinos.
If the mixing is large enough light sterile neutrinos will become thermalized and have the same temperature as active neutrinos before decoupling. 
This implies that they would contribute to $\Neff$. 
The modification to $\Neff$ due to light sterile neutrinos depends on the mass and mixing with the active states. 
This is shown in Figure~\ref{fig:n_effective} (left) for a mass-square-difference compatible with LSND and MiniBooNE observations.
Decreasing the mass compared to the scenario shown in this figure reduces the value of $\Neff$, while increasing it has the opposite effect as discussed in~\cite{Gariazzo:2019gyi}.
Recent constraints on $\Neff$ from~\cite{Pitrou:2018cgg} are shown in~ Figure \ref{fig:n_effective} (right) and indicate a tension between the measurements of $\Neff$ and the expectation from light sterile neutrinos.
These have been translated in terms of electron- and muon-neutrino mixing amplitudes and mass-squared-difference in~\cite{Hagstotz:2020ukm} and are shown in Figure~\ref{fig:cosmo_constraints}.
As can be observed from this figure, cosmological constraints are significantly stronger over most of the relevant parameter space both in electron-neutrino and muon-neutrino channels, except for regions of relatively small mass-square differences where they are comparable to muon-neutrino disappearance constraints; however, these regions are not preferred by recent global fits which favor $\Delta m^2 > \SI{1}{eV}^2$~\cite{Diaz:2019fwt}.

Similarly to the situation with terrestrial probes of light sterile neutrinos, the tension with cosmological observations~\cite{Hannestad:2012ky,Lattanzi:2017ubx,Knee:2018rvj,Berryman:2019nvr,Gariazzo:2019gyi,Hagstotz:2020ukm,Adams:2020nue} has propelled the search for possible solutions~\cite{PhysRevD.52.6595,PhysRevLett.93.081302,PhysRevD.87.023505,Hannestad:2013ana,Archidiacono:2014nda,Archidiacono:2015oma,Dasgupta:2013zpn,Saviano:2014esa,Chu:2015ipa,Chu:2018gxk,Song:2018zyl}.
The proposed solution can be organized in three main categories: large chemical potential or asymmetries in the active neutrino flavors~\cite{Foot:1995bm,Abazajian:2004aj,Mirizzi:2012we,Saviano:2013ktj}, secret neutrino interactions~\cite{Dasgupta:2013zpn,Hannestad:2013ana,Saviano:2014esa,Archidiacono:2014nda,Archidiacono:2016kkh,Chu:2015ipa,Forastieri:2017oma,Chu:2018gxk,Farzan:2019yvo,Cline:2019seo,Cherry:2016jol}, and modifications on the universe thermal history~\cite{Gelmini:2004ah,Gelmini:2019esj,Gelmini:2019wfp,Yaguna:2007wi,Hasegawa:2020ctq}

\section{Gravitational waves as light relics}

While additional particle species may exist and contribute to $N_\mathrm{eff}$ as described
previously, one source of radiation beyond the Standard Model is already present in Einstein's
general relativity: gravitational waves.
Formally, gravitational waves (GWs) backreact onto the expansion of the universe at second order in
perturbation theory, contributing an effective energy density and
pressure~\cite{Abramo:1997hu,Brandenberger:2018fte,Clarke:2020bil}.
GWs with wavelengths much smaller than the Hubble radius behave like a relativistic, spin-$2$ degree
of freedom; their energy density is therefore constrained by measurements of $N_{\rm eff}$.

The contribution of gravitational waves to the effective number of relativistic degrees of freedom  is~\cite{Maggiore:1999vm}
\begin{equation}
\label{eq:defNeff}
    \Delta N_{\rm eff, GW}
= \frac{8}{7} \left(\frac{11}{4}\right)^{4/3}
        \frac{\Omega_{\mathrm{GW}, 0}}{\Omega_{\gamma, 0}}
    = \frac{\Omega_{\rm GW, 0} h^2}{5.61 \times 10^{-6}}
\end{equation}
where $\Omega_{\gamma}$ and $\Omega_{\rm GW}$ are the critical fraction of standard model radiation and GWs, respectively, and a subscript $0$ denotes their value today.
The abundance of GWs (relative to the energy density in a flat universe, $\rho_\mathrm{crit}$) is
an integral over the contribution from all subhorizon wavelengths,
\begin{equation}\label{eq:GWint}
\Omega_{\rm GW}
= \frac{1}{\rho_{\rm{crit}}} \int_{\ln f_\star}^{\infty} d\ln f\frac{d\rho_{\rm GW}(f)}{d\ln f}
\end{equation}
The lower bound on the integral depends on the probe in question, i.e., it is some frequency smaller
than the horizon scale at the time of nucleosynthesis or smaller than the scales observed in the
CMB.
Constraints on $\Delta N_{\rm eff}$ thus provide a powerful indirect probe of GW backgrounds across
a wide frequency range, extending beyond the reach of current or planned GW detectors.
Note that constraints on the integrated abundance of GWs roughly correspond to a constraint on the
peak amplitude of the spectral abundance $\Omega_\mathrm{GW}(f)$, provided the peak is not narrower
than an $e$-fold in frequency.

Despite the lack of frequency-dependent information, the utility of a global upper bound on
$\Omega_\mathrm{GW}$ is particularly pronounced for stochastic backgrounds at present-day frequencies
in the range of $\mathrm{kHz}$ to $\mathrm{GHz}$.
The signal-to-noise ratio for direct detection experiments depends not upon the energy density in
GWs, but on their power spectral density $S_h(f) \propto \Omega_\mathrm{GW}(f) / f^3$.
Measurements of $N_\mathrm{eff}$ thus act as a probe of high-frequency stochastic GW backgrounds
with extremely favorable frequency scaling compared to direct detectors, which in the near term will
not be competitive for frequencies beyond $\mathrm{kHz}$ (the upper reach of ground-based
interferometry).
See Ref.~\cite{Aggarwal:2020olq} for a review of sources and detectors of ultra-high frequency GWs.

In standard $\Lambda$CDM cosmology, 
the relative precision of measurements of $\Delta N_\mathrm{eff}$ translates closely to the
corresponding abundance of gravitational waves in the early universe: applying conservation of
entropy,
\begin{equation}
    \frac{\Omega_\mathrm{GW}(a)}{\Omega_{\mathrm{rel}, \mathrm{SM}}(a)}
    \approx 1.06
        \cdot
        \frac{106.75}{g_{\star}(a)}
        \left( \frac{106.75}{g_{\star S}(a)} \right)^{-4/3}
        \frac{\Delta N_{\mathrm{eff}} }{ N_{\mathrm{eff}} }.
\end{equation}
The abundance of relativistic SM species $\Omega_{\mathrm{rel}, \mathrm{SM}}(a)$ is unity in the
early universe unless GWs (or other relics that were never in equilibrium with the SM) contribute
non-negligibly.
Current limits of $\Delta N_\mathrm{eff} \sim 0.2$ from Big Bang
Nucleosynthesis (BBN)~\cite{Cyburt:2015mya} and {\it Planck} data~\cite{Cyburt:2015mya} thus limit the abundance
of GWs to $\lesssim 7\%$ in the early universe, which next-generation CMB experiments may improve by
upwards of an order of magnitude~\cite{Pagano:2015hma, Abazajian:2019eic}.
We will now describe how these $\Delta N_{\rm eff}$-derived limits constrain
cosmological sources of GW backgrounds and the expansion and thermal history of the universe.

\subsection{Primordial gravitational-wave background from inflation}

A stochastic gravitational-wave background (SGWB) from primordial tensor fluctuations is generically produced in the inflationary paradigm~(cf.~the dedicated Snowmass~2021 White Paper~\cite{Snowmass2021:Inflation}). In fact, even before inflation was proposed, it was realized that the expansion of the universe allows for amplification of gravitational waves and production of gravitons~\cite{1974ZhETF..67..825G, 1977NYASA.302..439G}. Such a process of \emph{parametric amplification} requires that (1) modes spend time outside the Hubble radius (i.e., the background universe expands more rapidly than GWs vary in time), when (2) the universe is not radiation-dominated (RD).

Inflation naturally meets both requirements above
and hence enables production of macroscopic GWs from initial quantum fluctuations of the vaccum. After inflation ends, tensor modes start to reenter the Hubble radius 
and each, thereafter, redshifts like radiation. 
Together, all modes that reentered and remain inside the Hubble radius constitute the primordial SGWB.
For a given mode, the parametric amplification regime spans its horizon exit and reentry, 
and the amplification coefficient is just the ratio of the scale factors at these horizon crossings
\cite{1993CQGra..10.2449G}.

\paragraph{Stiff amplification of the primordial SGWB}
While all modes of interest exit during inflation, 
different modes can reenter during post-inflationary eras with different equations of state. For nearly scale-invariant superhorizon amplitudes, the contribution to 
$\Omega_{\rm GW}(f)\equiv {\rm d}\,\Omega_{\rm GW}/{\rm d}\ln f$
by tensor modes that reentered during the radiation-dominated era is nearly frequency-independent.   
This results in a long ``plateau'' in $\Omega_{\rm GW}(f)$ today~\cite{1997PhRvD..55..435T}.
However, the equation of state of the universe before BBN are poorly constrained and may depart from $w=1/3$. 
If there is a pre-BBN era with an equation of state stiffer than radiation (i.e., $w>1/3$),
high-frequency modes that reenter during such an era will have their $\Omega_{\rm GW}(f)$ amplified 
\emph{relative to the plateau}. 
This effect is referred to as \emph{stiff amplification}~\cite{2017PhRvD..96f3505L,2021JCAP...10..024L}.
Stiff-amplified primoridal SGWB may significantly contribute to $\Delta N_\mathrm{eff}$,
and thus measurements of the latter provide a probe into nonstandard expansion histories in the early universe;
see also~\cite{1998PhRvD..58h3504G,1999CQGra..16.2905G,2008PhRvD..77f3504B,2011PhRvD..84l3513K,Figueroa:2019paj}.
Furthermore, since a stiff era can arise from the early kination phase of scalar field dark matter (either real or complex),
during which $w_\mathrm{SF}=1$~\cite{2014PhRvD..89h3536L},
any evidence of a $\Delta N_\mathrm{eff}$ from amplified primordial GWs may even shed light on the nature of dark matter.

\subsection{Additional sources of cosmological gravitational-wave backgrounds}

Stochastic GW backgrounds can be sourced by a number of subhorizon nonlinear processes in
the early universe.
As the universe expands and cools, a variety of scenarios predict a brief phase characterized
complex, nonlinear field dynamics, including phase transitions, nonperturbative particle production,
and the formation of solitons or other topological defects (see, e.g.,
Refs.~\cite{Amin:2014eta,Allahverdi:2020bys}).
Such processes generate anisotropic stress which sources a GW background often characterized by a
narrow peak cut off at some peak frequency and a power-law tail at lower
frequencies~\cite{Hook:2020phx}.
The signal today would be observed at frequencies $f_0 \sim \sqrt{H_0 H_\mathrm{g}}$ in terms of the
Hubble rate at the time of generation, $H_\mathrm{g}$~\cite{Easther:2006gt,Dufaux:2007pt,Amin:2018kkg}.
The peak amplitude of the GW signal when it was generated (at a scale factor $a_g$) approximately
scales as~\cite{Easther:2006gt,Dufaux:2007pt,Amin:2014eta,Giblin:2014gra}
\begin{equation}
    \Omega_{\rm GW}(a_g, k_\mathrm{peak})
    \sim \left( \frac{a_{\rm g} H_{\rm g}}{k_\mathrm{peak}}\right)^2 \delta_{\pi}^2\,,
\end{equation}
where $\delta_{\pi}$ is the fraction of the total energy budget of the universe that sources GWs and
is bounded to be $\delta_{\pi}\lesssim 1$ in the case of a maximally inhomogeneous
universe. 
Furthermore, such nonlinear processes occur inside the horizon, so
$a_{\rm g} H_{\rm g} / k_\mathrm{peak} \lesssim 1$.
As such, in order for $\Delta N_\mathrm{eff}$ bounds on the integrated spectrum to be useful, GWs
must be sourced on scales just inside the horizon or over a broader range (in $e$-folds) of
frequency.

Preheating, a phase of resonant particle production at the beginning of reheating after inflation, and associated inhomogeneous dynamics 
can lead to $\delta_\pi$ which is not too small compared to unity.
In many scenarios involving scalar fields, particle production occurs on scales well within the
horizon and often does not deplete the entirety of the inflaton's energy~\cite{Kofman:1997yn},
leading to a signal with peak amplitude
$\lesssim\mathcal{O}(10^{-9})$~\cite{Easther:2006gt,Dufaux:2007pt,Amin:2014eta,Giblin:2014gra}.
In contrast, gauge-field--inflaton couplings offer an especially efficient preheating channel:
strong tachyonic amplification of the gauge fields on scales just inside the horizon, draining
nearly $100\%$ of the inflaton's energy within $2$ $e$-folds of expansion~\cite{Adshead:2015pva}.
For axion-driven inflation, the mechanism is so efficient that existing $N_\mathrm{eff}$ constraints
already place constraints on the coupling to gauge
fields~\cite{Adshead:2018doq,Adshead:2019lbr,Adshead:2019igv}, while future observations will have
sufficient sensitivity to rule out the model as a preheating mechanism.
In hybrid inflation models, or if the inflationary energy scale is sufficiently low, the signal
could also be visible at interferometers~\cite{Cui:2021are}.

Other scenarios which can give rise to substantial GW signals, of interest for future
direct-detection experiments and cosmological probes of $\Delta N_{\rm eff}$, involve first order
phase transitions and cosmic strings.
Phase transitions source GWs due to subhorizon nonlinear processes like bubble collisions, sound
waves produced by the bubble walls, and the turbulence following the completion of the phase
transition (for a review, see, e.g.,~\cite{Hindmarsh:2020hop}).
Phase transitions in gauge theories can also lead to the formation of cosmic string networks.
Their evolution can be described with a scaling solution for which the total density of the strings is a constant small fraction of the total density.
As the universe expands, on near-horizon scales the string network is broken down into loops, which undergo multiple oscillations and eventually decay into GWs.
This continuous emission of GWs gives rise to a characteristic plateau in the GW power spectrum for
a standard expansion history before BBN and can be sensitive to deviations from it (for a recent review, see, e.g., Ref.~\cite{Auclair:2019wcv}).

\paragraph{Amplification of signals in nonstandard expansion histories}

If any of the above mechanisms generated gravitational wave before the universe reheated (i.e.,
before the Standard Model was produced and reached equilibrium), the expansion of the universe in
the intervening period modulates $\Omega_{\rm GW,0}$.
In particular, a stiff equation of state $w > 1/3$ can boost the GW signal significantly even for a
modest number of $e$-folds of expansion between GW generation and the end of reheating.

\subsection{Observational prospects}

As mentioned before, cosmological GW backgrounds can be searched by indirect probes, 
e.g., light element abundances from BBN, the CMB, and large-scale structure of the universe. 
These probes measure the value of $N_\mathrm{eff}$
and provide what is known as ``integral bounds'' on the SGWB; see Eq.~(\ref{eq:GWint}). It should be noted that $N_\mathrm{eff}$ at different epochs can be different, 
not only because of different lower bounds on the $\Omega_\mathrm{GW}$ integral 
but because components other than GWs may additionally contribute to $N_\mathrm{eff}$ at a given epoch.
For example, an early stiff matter may boost the expansion rate during BBN,
enhancing the value of $N_\mathrm{eff}$ then~\cite{2017PhRvD..96f3505L}.
Therefore, caution must be exercised when combining $N_\mathrm{eff}$ measured by different probes.
Another interesting aspect of GWs as radiation relics comes from 
the possibility to alleviate the Hubble tension with non-negligible $\Delta N_\mathrm{eff}$~\cite{2019JCAP...10..029S}. 
Currently, the high-$z$ measurements of the Hubble constant from the CMB + BAO data can admit a higher value of $H_0$, 
closer to those from local distance-ladder measurements, 
if fitted by the seven-parameter $\Lambda$CDM + $N_\mathrm{eff}$ model. 
A cosmological SGWB then naturally provides such a contribution to $N_\mathrm{eff}$ 
without introducing other exotic components or new physics~\cite{2021JCAP...10..024L}.

Apart from the indirect probes, there are also direct probes of relic GW backgrounds across a wide range of frequencies~\cite{Lasky:2015lej}.
Existing approaches include the CMB temperature and polarization, pulsar-timing array (PTA), and GW laser interferometry, 
from low to high frequencies.

The primordial SGWB can leave an observable imprint on the CMB temperature and polarization anisotropy (e.g.,~\cite{1993PhRvL..71..324C}).
In particular, detection of the CMB $B$-mode polarization around $\ell\sim 100$ 
would be a convincing signature of the primordial SGWB.
Recently, BICEP3/Keck Array has updated the upper bound on the tensor-to-scalar ratio,
$r_{0.05}<0.036$ at $95\%$ confidence~\cite{BICEP:2021xfz}.
In the future, CMB-S4 will continuously seek to measure the primordial SGWB from inflation~\cite{CMB-S4:2016ple,Abazajian:2019eic}.
For further details, we refer to the~\cite{Snowmass2021:Inflation}.

PTA observations measure the times of arrival (“ToAs”) of radio pulses from millisecond
pulsars. Those ToAs can be modulated by an SGWB permeating the spacetime between
the pulsar and the earth.  Recently, NANOGrav reported strong evidence for a stochastic common-spectrum process around $f_{\rm yr}=1$~yr$^{-1}$
in their 12.5~yr pulsar-timing data set~\cite{NANOGrav:2020bcs}, and such a process was also identified in the PPTA data later~\cite{Goncharov:2021oub}.
Though it has not been confirmed as an SGWB detection due to a possible misspecification of the model 
and the lack of spatial correlations in the signal, 
many interpretations in this direction have flourished since then,
including the primordial SGWB with a large initial blue tilt from non-standard inflation~\cite{2021MNRAS.502L..11V,2021JCAP...01..071K} 
and the GWs from a first-order phase transition~\cite{Xue:2021gyq,NANOGrav:2021flc}.
On the other hand, the primordial SGWB from \emph{standard} inflation 
cannot account for such a common spectrum process even with stiff amplification~\cite{2021JCAP...10..024L}.

Laser interferometers like the Advanced LIGO-Virgo network can directly detect SGWBs 
by cross-correlating data from different detectors (e.g.,~\cite{2017LRR....20....2R}).
Recently, the LIGO Scientific Collaboration and Virgo Collaboration published results of a search for an isotropic SGWB
using data from their first three observing runs (O1, O2 and O3)~\cite{KAGRA:2021kbb}.
While the cross-correlation spectrum from data does not show evidence for an SGWB signal, 
a new upper limit is placed on the present-day SGWB energy spectrum, 
modeled as a power law around $f_{\rm ref}=25$~Hz. This upper limit is expected to be continually improved by future upgrades of LIGO, e.g.,~\cite{LIGOA+}.
In the meantime, other commissioned interferometer experiments like the LISA-Taiji network and TianQin
will provide detection channels of cosmological SGWB at different frequency bands,
and future GW detectors are proposed to fill the gaps in the frequency spectrum~\cite{Sesana:2019vho,Aggarwal:2020olq}.

\section{Conclusion}
New light species arise in many well-motivated extensions of the Standard Model.  Any new light states produced in the early universe will leave observable imprints through their gravitational influence on the expansion history and evolution of density fluctuations.  Upcoming cosmological observations will greatly improve the precision with which we measure the energy density and properties of light relics, providing broad insights into physics beyond the Standard Model.

Light relics that were in thermal equilibrium with the Standard Model plasma in the early universe make a contribution to the light relic density that can be calculated from their spin and freeze-out temperature.  Each species of light thermal relic contributes $\Delta \Neff \geq 0.027$, thereby establishing clear thresholds that help to elucidate the value of precise measurements of the cosmological radiation density.  Late thermalization of light species can produce even larger contributions to $\Neff$, and upper bounds on $\Delta \Neff$ could be used to rule out various types of DM-baryon interactions.

Free-streaming light relics create a characteristic shift to the phase of acoustic oscillations in the photon-baryon plasma. This effect is absent for light relics that do not free-stream due to self-interactions or scattering with other species.  Observations of the CMB and LSS can therefore distinguish between free-streaming and fluid-like light relics.

Light relics need not be exactly massless, and non-zero masses can lead to additional signatures in cosmological observables.  Light species that become non-relativistic after decoupling act as hot dark matter, suppressing the growth of structure on scales smaller than their free-streaming scale, an effect that can be measured with CMB lensing and LSS surveys.  

The dark sector may involve a rich set of particle types and interactions, much like the visible sector described by the Standard Model.  This richness may be motivated by models aimed at addressing challenges in particle physics, like the hierarchy problem, or puzzles in cosmology, like the Hubble tension.  However, complexity in the dark sector often involves new light states, and cosmological measurements of the light relic density thereby constrain the range of allowed possibilities.  

Anomalies in short baseline neutrino oscillation experiments have motivated models containing new sterile neutrino states. Since sterile neutrinos contribute to the light relic density, cosmological measurements of $\Neff$ place broad constraints on the range of viable sterile neutrino models.

Gravitational waves present in the early universe contribute to the light relic density.  A stochastic gravitational wave background that makes a significant contribution to the radiation density may arise due to  preheating, phase transitions, topological defects, or the amplification of inflationary gravitational waves.  Cosmological measurements of the light-relic density serve as an integral constraint on the stochastic gravitational wave background, thereby placing constraints on these scenarios.

Upcoming cosmological observations will either detect new contributions to the light relic density, or place severe constraints on the extensions to the Standard Model that involve new light species.  In either case, this will have broad implications for models of physics beyond the Standard Model.

\vspace{0.5cm}
\subsection*{Acknowledgments} 

CD is partially supported by US~Department of Energy~(DOE) grant~\mbox{DE-SC0020223}.
JM is supported by the US~Department of Energy under Grant~\mbox{DE-SC0010129}.
TC is supported by the US~Department of Energy, under grant number~\mbox{DE-SC0011640}.
NC is supported by the US~Department of Energy under grant number~\mbox{DE-SC0011702}.
PD is supported in part by Simons Investigator in Physics Award~623940 and NSF award~\mbox{PHY-1915093}.
JBM is supported by a Clay fellowship at the Smithsonian Astrophysical Observatory. 
KS is supported by a Natural Sciences and Engineering Research Council of Canada~(NSERC) Subatomic Physics Discovery Grant. 
BW was supported by the US~Department of Energy under Grants~\mbox{DE-SC0009919} and~\mbox{DE-SC0019035}, and the Simons Foundation under Grant~SFARI~560536. WLX is supported by the US-Department of Energy under Contract~\mbox{DE-AC02-05CH11231}.

\bibliographystyle{utphys}
\bibliography{light_relics.bib}

\end{document}